\documentclass[12pt, a4paper]{article}

\usepackage[utf8]{inputenc}
\usepackage{amsmath}
\usepackage{amssymb}
\usepackage{bm}
\usepackage{bbold}
\usepackage{mathrsfs}
\usepackage{float}
\usepackage{setspace}
\usepackage{graphicx}

\usepackage{hyperref}
\hypersetup{colorlinks =true,allcolors=blue}

\usepackage{csvsimple} 
\usepackage{titling}
\usepackage{authblk} 

\usepackage[lmargin=25mm,rmargin=25mm,tmargin=35mm, bmargin=35mm]{geometry}
\usepackage{lineno}
\usepackage{lipsum}

\usepackage[backend=bibtex,  natbib=true, style=authoryear, maxbibnames=2, maxnames=2,
            doi=false,isbn=false, url=false, eprint=false]{biblatex}
\usepackage{biblatex}

\renewcommand{\vec}[1]{\mathbf{#1}}

\title{Borrowing from historical control data in a Bayesian time-to-event model with flexible baseline hazard function}

\author[1]{DARREN A. V. SCOTT}
\author[2]{ALEX LEWIN}
\affil[1]{AstraZeneca R\&D, Cambridge Darren.Scott@astrazeneca.com}
\affil[2]{Department of Medical Statistics, London School of Hygiene and Tropical Medicine, Keppel Street, London,
	UK, WC1E 7HT}

\date{}                     
\begin{document}

\maketitle

\section*{Abstract}

Currently, there is a focus on statistical methods that can use historical trial information to help accelerate the discovery, development, and delivery of medicines. Bayesian methods can be constructed so that the borrowing is ``dynamic", the similarity of the data helps determine how much information is used. In the time-to-event setting, Bayesian methods typically impose a constraint on the shape of the baseline hazard. We propose a Bayesian borrowing semiparametric model for one historical dataset, which allows the baseline hazard to take any form through an ensemble average. We introduce
priors to smooth the shape of the posterior baseline hazard, improving both model estimation and borrowing characteristics. We explore a variety of prior structures for the borrowing within our proposed model and assess their performance against established approaches. We show the benefit of a lump-and-smear borrowing priors in our joint model for improving the type I error in the presence of prior-data conflict and increased power. A principled approach is proposed for the choice of hyperparameters to control the dynamic borrowing, based on the tolerated difference between the historical and log baseline hazards. We have developed accompanying software available in R enabling easy implementation of our approach.   

\emph{Key words:} Bayesian borrowing, mixture priors, Gaussian Markov random field prior, commensurate prior, time-to-event.

\newpage

\section{Introduction}

%



Traditionally, randomised clinical trials are designed and analysed from a frequentist perspective using classical hypothesis testing. However, more recently, there has been a growing awareness of the benefits of Bayesian approaches, which naturally allow for the explicit integration of previous knowledge with new empirical data. This is particularly attractive in clinical trials, where multiple trials are often conducted on the same illness in the pursuit of an effective treatment or an existing therapy has been established. 

Borrowing information from an appropriate historical trial is appealing to practitioners and regulators for reasons of improved efficiency from smaller and faster trials, increased statistical power, and fewer patients assigned to a less attractive treatment. It has been used extensively in medical device trials and is increasingly being used in drug trials in oncology \citep{Su2022} and pediatrics \citep{Savic2017}.  There are several distinct statistical methods to do this including test-then-pool \citep{Viele2014}, commensurate priors \citep{Hobbs2011}, hierarchical models \citep{Neuenschwander2010}, power priors \citep{Ibrahim2015} and meta-analytic-predictive priors \citep{Schmidli2014}.  

One challenge confronted by researchers hoping to design a trial with retrospective data is understanding the commensurability of the information with the trial data that has yet to be collected. Questions concerning the comparability of the patient population, the design of the trial, and the standard of care that induces the ``placebo" effect should be considered. If the historical information differs substantially from the new current trial,  borrowing from a biased source leads to an inflated Type I error rate, as well as the possibility of needing to run a longer, more expensive trial in order to overcome the incommensurate prior data. The benefit of Bayesian borrowing is that it is ``dynamic", the amount of historical data borrowed is related to the agreement between the datasets. 

We are concerned with leveraging information from a historical control for time-to-event endpoints such as time-to-disease progression or time-to-death within a model where the hazard function is free to take a range of shapes. These endpoints are the primary outcome in a variety of therapeutic areas, including oncology
and cardiovascular disease. \cite{Hobbs2013} propose a piecewise exponential model (PEM) with a commensurate prior to control the level of borrowing. They assume predetermined fixed intervals and independence across the associated baseline hazards. \cite{Han2017} use the PEM with fixed intervals and independent baseline hazards over time, to borrow the control effect across multiple studies. They incorporate patient-level covariates to enhance the efficiency of borrowing. \cite{Bi2023} model the baseline hazard with an exponential distribution and use a Dirichlet process to cluster the historical baseline hazards. This helps to discount the historical data in the presence of prior-data conflict. \cite{Lewis2019} use Weibull likelihood with a simple commensurate prior in a joint model to borrow form a single historical dataset. Finally, \cite{Roychoudhury2020} extend the PEM to multiple datasets using a hierarchical structure and a robust meta-analytic predictive prior. All of these approaches impose a constraint on the shape of the baseline hazard over time; either as a horizontal line, a step function, or a monotonic function.

We use a joint semiparametric hierarchical model to borrow information from a historical
time-to-event dataset with a smooth and flexible baseline hazard
time-to-event dataset. By accurately capturing the true underling hazard function, we are to improve the borrowing characteristics of the model (power and type I error in the presence of prior-data conflict), compared with approaches which constrain the hazard function. We use an ensemble method, popular in tree-based approaches such as random forests, to obtain a flexible baseline hazard function. To smooth the shape of the baseline hazard function, we introduce a dependency across the time points between the log baseline hazards via a Gaussain Markov random field (GMRF) prior, improving the accuracy of our estimation. 

We interpret the commensurate priors, which control the degree of borrowing in the model, from two perspectives; in terms of prior information on the current log baseline hazard and by the characteristics of the dynamic borrowing
component. We show how lump-and-smear priors can improve the characteristics of dynamic borrowing, making the borrowing robust to violations of the assumptions of exchangeability and prior-data conflict. We recommend a procedure for selecting their values based on a tolerance for differences between historical and current log baseline hazards.

Our proposed semiparametric Bayesian borrowing model requires individual participant data (IPD),  rather than summary data, as is the case with some other recent approaches (\cite{Roychoudhury2020}, \cite{Bi2023}). Although this is clearly more convenient from a logistical perspective, there is a trade-off for this convenience. We use the additional information from the IPD for a Bayesian borrowing model which has a baseline hazard which can take any form. With summary-level data, the shape of the baseline hazard is constrained, either to a fixed value \cite{Bi2023} or to a step function \cite{Roychoudhury2020}. By having a model which parameterises the baseline hazard to allow borrowing but is free to accurately capture the underlying baseline hazard,  which we typically expect to be a smooth function of time, we are able to improve the power when borrowing is appropriate and reduce the potential for type I error in the presence of prior-data conflict. The IPD also allows for covariates (which do not have to be the same across the datasets) to be incorporated into both the historic and current trial, which can help reduce posterior uncertainty within our model. Again, this is not applicable for the summary-level data models. 

The primary aim of our paper is to provide a unified framework for flexible, fully Bayesian analyses of time-to-event data which dynamically borrows from a historical control, along with user-friendly software (available in the Comprehensive R Archive Network (CRAN)) that enables investigators to use our methods in a variety of settings. We develop a reversible-jump Monte Carlo Markov chain (RJMCMC) algorithm and software which utilises the power prior and conjugacy to improve the sampling properties and avoid unnecessary tuning parameters. 

We explore how the prior structure and the choice of smoothing parameters impact the posterior estimate of the historical baseline hazard. We propose three different types of prior for the borrowing of the log baseline hazard in the presence of between-trial heterogeneity. We compare our model with the different borrowing priors with two approaches, which are commonly used in a time-to-event setting. Finally, we demonstrate the approach by applying our method in a reanalysis of a randomised control trial for the treatment of asthma, by borrowing external placebo patients from a historical trial within our proposed approach.


\section{Methods}
In this section, we discuss our approach to leveraging an historical dataset for a time-to-event outcome when the assumption of proportional hazards holds. We consider the setting with a control and treatment arm, for which the use of historical data is confined to the control group. The prior distribution for the control baseline hazard will be informed by historical data. The prior for the treatment contrast will be vague. 

We wish to have a model where the baseline hazard can take any shape, rather than a monotonic function or a simple step function. To achieve this, we use an ensemble method, popular in tree-based regressions such as random forests, to obtain the posterior baseline hazard function. For each iteration, we partition time into intervals using split points and sample the posterior baseline hazards for each interval. We recursively alter the partition region (and therefore posterior baseline hazard) by changing the location of the split points or adding or removing split points (or both). The expected posterior baseline hazard is a smoothed function obtained via an ensemble average. Because the baseline hazard function can be very flexible, we use our prior structure to smooth its shape.   

There are three key challenges to setting up the Bayesian model; the specification of the hazard function, the choice of prior distribution, and the development of a robust and efficient computational scheme. In this section, we describe our solutions to these challenges. 
  
\subsection{Bayesian semiparametric time to event model}
In order to obtain a flexible baseline hazard, we begin with a PEM that can be expressed in terms of the hazard for random variable of a time of event $Y$ as $h(Y) = \lambda_{j}$ where $j$ is the interval containing $Y$. This leads to a likelihood of the data 
\begin{equation}
    \mathcal{L}(\vec{y}; \bm{\lambda}) = \prod_{i=1}^n h(y_i)\exp\left(-\sum_{j} \lambda_jE_{ij} \right),
\end{equation}
where $E_{ij}$ is the exposure time for the $i$th individual within the $j$th time interval ($y_i = \sum_j E_{ij})$. 

So that each ensemble has a different partition of time, we make the location of the split points and the number of the split points parameters in our model. This allows us to average over the various partitions of time to obtain a smooth flexible hazard function. The approach is analogous to Bayesian model averaging, where we integrate over all possible piecewise exponential models weighted by their respective probabilities. Smoothing of the hazard function is also assisted by our prior structure of the baseline hazard.   

More specifically, to incorporate covariates into our ensemble method, the likelihood of the current trial consists of $J+1$ intervals with split points $0=s_0 <s_1 <...<s_{J+1}$ with $s_{J+1} > y_i$ for all $i=1,2,...,n$. Thus, we have $J$ partitions and $J+1$ intervals $(0, s_1],...(s_{J-1}, s_{J+1}]$. In the $j$th interval, we assume a constant baseline hazard $h_{j}(y_i) = \lambda_{j}$ for $y_i \in I_j = (s_j-1,s_j]$· In order to ensure that the algorithm is stable $s_{J+1}$ must be set to $\text{max}(y_i|\nu_i =1)$, the longest non-censoring time observed.

Let $\vec{D} =(n,\vec{y},\vec{X},\bm{\nu})$ denote the observed data with $\nu_i=1$ if the $i$th subject failed (death or event happens) and 0 otherwise, and $\vec{X}$ is the $n \times p$ matrix
of covariates with $i$th row $\vec{x}'_i$. The likelihood function can be expressed 
\begin{align}
    \label{eq:conc_like}
    \mathcal{L}(\vec{D};\bm{\beta}, \bm{\lambda}, \vec{s},J) =
           & \prod_{i=1}^n\prod_{j=1}^{J+1} (\lambda_j \exp\left\{\vec{x}_i'\bm{\beta}\right\})^{\delta_{ij}\nu_i}
     \nonumber \exp \Big\{
    -\delta_{ij} \Big(
    \lambda_j(y_i-s_{j-1})\\
    & \qquad  + \sum_{g=1}^{j-1}\lambda_g (s_g-s_{g-1})
    \Big)\exp\{\vec{x}_i'\bm{\beta}
    \}
    \Big\}
\end{align}

where $\delta_{ij} = 1$ if the $i$th subject failed or was censored in the $j$th interval, and
0 otherwise, $\bm{\lambda} = (\lambda_1, ..., \lambda_J)$ is the vector of baseline hazards, $\vec{x}_i' = (x_{i1} , x_{i2}, ... , x_{ip})$ denotes the $p \times 1$ vector of covariates
for the $i$th subject, and $\bm{\beta} = (\beta_1, ..., \beta_p)$ is the corresponding vector of regression coefficients.

The likelihood of the historical data takes the same functional form, with different parameters and data but the same number and placement of the split points $\vec{s}$,  $\mathcal{L}(\vec{D}_0;\bm{\beta}_0, \bm{\lambda}_0, \vec{s}, J )$ where $\vec{D}_0=(n_0,\vec{y}_0,\vec{X}_0,\bm{\nu}_0)$. By making the placement of the intervals $\vec{s}$ and the number of intervals $J$ parameters in the model, they are free to vary according to the information in both historical and current data. This means that each ensemble will comprise different partitions of time. The baseline hazard will be constant throughout each interval, but as the partitions change, our ensemble average will be able to capture the underlying shape of the hazard function. 

We allow conditioning on baseline covariates, as this can also help to reduce the standard error of the treatment effect and increase the power. This has been done in the case of a fixed baseline hazard by \cite{Han2017}, where borrowing is on the parameters associated with the covariates. However, the model is not collapsible, as for all time-to-event models, so conditioning on the covariates changes the nature of the estimated treatment effect  \citep{Daniel2021}.

\subsection{Smoothing prior for baseline hazards}

Here, we describe the priors that we use to smooth the baseline hazard function from the ensemble approach. The priors for the time component where time intervals are defined by the split points $0=s_0 <s_1 <...<s_{J+1}$ in (\ref{eq:conc_like}) and $s_0$ and $s_{J+1}$ are fixed are  
\begin{align}
   J|\textit{\small{J}}_\text{max} &\sim  \text{Poisson}_\text{t}(\phi) \qquad 0 \leq J \leq J_\text{max}, \label{eq:sp_prior}\\
   \vec{s}|J & \propto (2J+1)! \prod_{j=1}^{J+1} \frac{s_j - s_{j-1}}{s_{J+1}^{2J+1}}.
\end{align}
The total number of split points $J$ is a random variable from a right-truncated Poisson distribution at $J_\text{max}$ with parameter $\phi$, where $\phi$ is fixed and determines our prior belief of the mean number of split points. The combination of $\phi$ and the truncation point $J_\text{max}$ can have a large effect on smoothing the posterior baseline hazard. The truncation at $J_\text{max}$ shrinks the mean number of split points, and to a greater extent the variance, below $\phi$. (The density and associated moments are in the Supplementary Material.) 

The distribution of $\vec{s}$ conditional on the number of splits is the even-numbered order statistics of $2J + 1$ points uniformly distributed over $[0, s_{J+1}]$ \citep{Green1995}. This strategy of using even-numbered order statistics is adopted to prevent the splits from being too close together, which helps avoid having intervals that contain only a few or no events.

The prior structure that links the baseline hazards (Section \ref{sec:commp}) implies a constraint on the location of the split points for datasets where the maximum time to event in the historical and current study differ. The $s_J$ split point must occur before the supremum of the maximum survival time, and thus the proposal for the split point swap and birth-or-death move in the sampler is adjusted accordingly.  

Our prior structure for the log baseline hazard is motivated by our understanding that this function is likely to be a smooth function over time. Given the partition of the time scale, the log-baseline hazard is unlikely to be independent of each other a priori. We view the components of the historical $\log(\bm{\lambda}_{0})$ via a one-dimensional spatial problem, and specify a Gaussian Markov random field prior with nearest-neighbour structure in the form $\log(\lambda_{0j})|\left\{\log(\lambda_{0k}), j \neq k \right\} \sim N(\nu_j, \sigma^2_j)$

\begin{equation}
    \label{eq:ICAR_cond}
    \nu_j = \mu + \sum _{k \neq j} W_{jk}(\log(\lambda_{0k})- \mu).
\end{equation}
The hyperparameter $\mu = \mathbb{E}(\log(\lambda_{0j}))$ represents the overall trend in the levels of the log hazard function and $W_{j(j-1)} = l_j$ and $W_{j(j+1)}= r_j$ are the influences of the left and right neighbours of $\log(\lambda_j)$ respectively. All other $W_{jk}$ where $k \notin \{j-1, j, j+1 \}$ are set to 0. If we set $r_j = 0$ and $l_j =1$ we get the random walk prior process (autoregressive order 1) used by \cite{Murray2016}, a special case of our more general prior.  

The weights are specified so that of the two neighbours of $\log(\lambda_{0j})$ the larger interval has the greatest influence. A hyperparameter $c_\lambda \in (0,1)$ is included to allow the user to specify a prior belief in the level of dependency between the nearest neighbours and acts as a smoothing parameter in combination with the approximate mean choice of split points $\phi$. The combination of weights and variance guarantees that the joint distribution is Gaussian \citep{Besag1995}, ensuring computational efficiencies.  

The conditional prior of (\ref{eq:ICAR_cond}) and the considered choice of $\nu_j$ and $\sigma^2_j$ leads to the multivariate joint Gaussian prior (details are provided in the Supplementary Material), where $\sigma^2_\lambda$ is an overall measure of the variation across the baseline hazards with mean $\mu$.   
\begin{equation}
\pi(\log(\bm{\lambda}_{0})|c_\lambda)  \sim   \mathcal{N}_{J+1}(\mu \vec{1}, \sigma^2_\lambda  \Sigma_s)
\end{equation}
with hyper priors
\begin{align*}
\sigma^{2}_\lambda & \sim \text{IGamma}(a_\sigma,b_\sigma)\\
\pi(\mu) &\propto 1. 
\end{align*}

\subsubsection{Choice of smoothing parameters}
By treating the split points in the likelihood as random, and integrating over the uncertainty, the underlying baseline hazard function can be accurately modelled. There is a trade-off between flexibility and variability, when deciding on the smoothing parameters of $\phi$ ($J_{\text{max}}$) and $c_\lambda$. Although the baseline hazard can take more unusual shapes when $\phi$ is large and $c_\lambda$ is low, this is usually unrealistic. We explore the settings of the smoothing parameters in the simulation study and advise on how to set these from the results. 

\subsection{Commensurate prior for borrowing from historical information }\label{sec:commp}

The borrowing from the smoothed historical baseline hazard is controlled by a commensurate prior, which assumes that the
expected value of the outcome in a historical control arm is a potentially biased representation of its counterpart in the 
current trial. The drift of $\delta_j= \log(\lambda_j)-\log(\lambda_{0j})$ is assumed to be normally distributed $\delta_j \sim N(0,\tau_j)$. The crucial parameter is $\tau_j$ which controls the commensurability (or variance)) between historical and current control for the interval $I_j$
\begin{equation}
    \log(\lambda_j) \sim \mathcal{N}(\log(\lambda_{0j}), \tau_j).
\end{equation}

We explore three different structures for our commensurate prior 
\begin{align}
    \tau_j^{(\text{uni})}  &\sim \text{IGamma}(a_\tau, b_\tau) \label{eq:tau_trad} \\
    \tau_j^{(\text{mix})} &\sim p_0 \text{IGamma}(a_\tau, b_\tau) + (1 - p_0) \text{IGamma}(c_\tau,d_\tau) \label{eq:tau_mix} \\
    \tau  &\sim p_0 \text{IGamma}(a_\tau, b_\tau) + (1 - p_0) \text{IGamma}(c_\tau,d_\tau) \label{eq:tau_all} 
\end{align}
where the hyperparameters are fixed. Typically $a_\tau$ and $c_\tau$ are set to 1, leaving a choice for $p_0$ and $d_\tau$.  

The inverse gamma prior of (\ref{eq:tau_trad}) imposes a simple assumption of exchangability between the baseline hazards for interval $I_j$. The prior in \eqref{eq:tau_mix} is the most flexible, as each current log hazard $\log(\lambda_j)$ has its own commensurate parameter $\tau_j$. Alternatively, in \eqref{eq:tau_all}, one parameter $\tau$ controls the commensurability of the current log hazards at each interval.  The mixture priors of (\ref{eq:tau_mix}) and (\ref{eq:tau_all}) can be viewed as lump-and-smear priors which robustify the prior against prior-data conflict. We explore this concept and discuss how the priors can be interpreted in the next Section, so that the appropriate values can be chosen for the hyperparameters.

We account for baseline characteristics, which are free to differ between the two studies. The improper priors are 
\begin{align*}
    \bm{\beta}_0  &\propto 1, \qquad
    \bm{\beta} \propto 1. 
\end{align*}

\subsubsection{Choice of hyperparameters for the commensurate prior}\label{sec:pa_hyper}
We can understand the impact of the hyperparameters for our commensurate prior on the amount of borrowing through the conditional posterior of $\tau_j$.  Higher values of $\tau_j$ discount historical information, whereas smaller values constrain the current baseline hazard close to the historical counterpart, pooling the information. We propose a simple approach, using the historical data set and acceptable log baseline hazard differences, to select appropriate hyperparameter values. 

Imposing a mixture on $\tau_j$ allows for a combination of densities, a narrow lump over small values of $\tau_j$, which encourages borrowing, and a wide smear with mass over larger values, reflecting the size of the possible drift given the scale of the log baseline hazard. The lump-and-smear prior (Figure 2 in the Supplementary Material) ``robustifies" the model by allowing the possibility of large differences (or non-exchangeable log hazards), which the univariate prior (\ref{eq:tau_trad}) considers implausible.  We can interpret the weight parameter $p_0$ within the
commensurate mixture prior, as discounting the information $n_0$ by $1 - p_0$. 

The conjugate update for the mixture prior is   
\begin{align}
    \label{eq:condpost}
    \tau_j^{(\text{mix})}|\vec{D},\vec{D}_0, \lambda_j, \lambda_{0j} \sim&  (q_j)  \text{IG}\left(\frac{1}{2} +a_\tau, \frac{(\log(\lambda_j)-\log(\lambda_{0j}))^2}{2} +b_\tau\right) +  \nonumber\\
    & (1-q_j) \text{IG} \left(\frac{1}{2}+c_\tau, \frac{(\log(\lambda_j)-\log(\lambda_{0j}))^2}{2} +d_\tau\right),
\end{align}
where the updated weights ($q_j$, $1-q_j$) are proportional to the product of the original weights and the marginal likelihood with respect to the prior
\begin{align}
w_{0j} &= \frac{ b_\tau^{a_\tau}\Gamma(a_\tau + \frac{1}{2})}{(\frac{1}{2}(\log(\lambda_j) - \log(\lambda_{0j}))^2)+b_\tau)^{\frac{1}{2}+a_\tau}\Gamma(a_\tau) },\\
w_{1j} & =  \frac{ d_\tau^{c_\tau}\Gamma(c_\tau + \frac{1}{2})}{(\frac{1}{2}(\log(\lambda_j) - \log(\lambda_{0j}))^2)+d_\tau)^{\frac{1}{2}+c_\tau}\Gamma(c_\tau) }
\end{align}
$w_j = p_{0}w_{0j} + p_{1}w_{1j}$, 
$q_j = \frac{p_0w_{0j}}{w_j}$ and $(1- q_j) = \frac{p_1w_{1j}}{w_j}$. This reveals the limitation of the univariate prior, choosing $a_\tau=1$ and setting $b_\tau =0.001$ to encourage borrowing, results in a conditional posterior relatively insensitive differences in the log baseline hazard. The data-driven scale update $\frac{1}{2}(\log(\lambda_j)-\log(\lambda_{0j}))^2$, is too slow in adjusting the scale for prior-data conflict. The mixture prior transfers the data-driven update from the change in scale to the posterior weight, making the borrowing much more responsive to differences in the log baseline hazards.  

Large values of the posterior mixture weight $q_j$ result in greater borrowing, while small values of $q_j$ lead to discounting of the historical data.  We explore the impact of our choice of hyperparameters $d_\tau$ and $p_0$ on $q_j$, by setting $a_\tau = c_\tau=1$ and $b_\tau << d_\tau$.  We assume for the following discussion that given the scale of the true underlying baseline hazard a $d_\tau \geq 1$ will lead to a sufficiently diffuse inverse gamma that the sampled commensurate parameter will prevent borrowing. Thus, we restrict our attention to SSEb values below 1.6, as the scale for either element of the mixture beyond this, would be large enough to restrict borrowing. Simplifying $q_j$  and treating it as a function of the squared error of the log baseline hazards  (SSEb = $(\log(\lambda_j)-\log(\lambda_{0j}))^2$) for the interval $j$ and hyperparameters gives 
\begin{align}
 q_j(\text{SSEb}|b_\tau, d_\tau, p_0) &= \frac{1}{1 + \frac{(1-p_0)}{p_0}\frac{d_\tau}{b_\tau} \left(
 \frac{\frac{1}{2}\text{SSEb}+b_\tau}{\frac{1}{2}\text{SSEb}+d_\tau}\right)^{\frac{3}{2}}
 }. \label{eq:qMSE}
\end{align}

\begin{figure}[H]
	\centering
	\includegraphics[width=14cm]{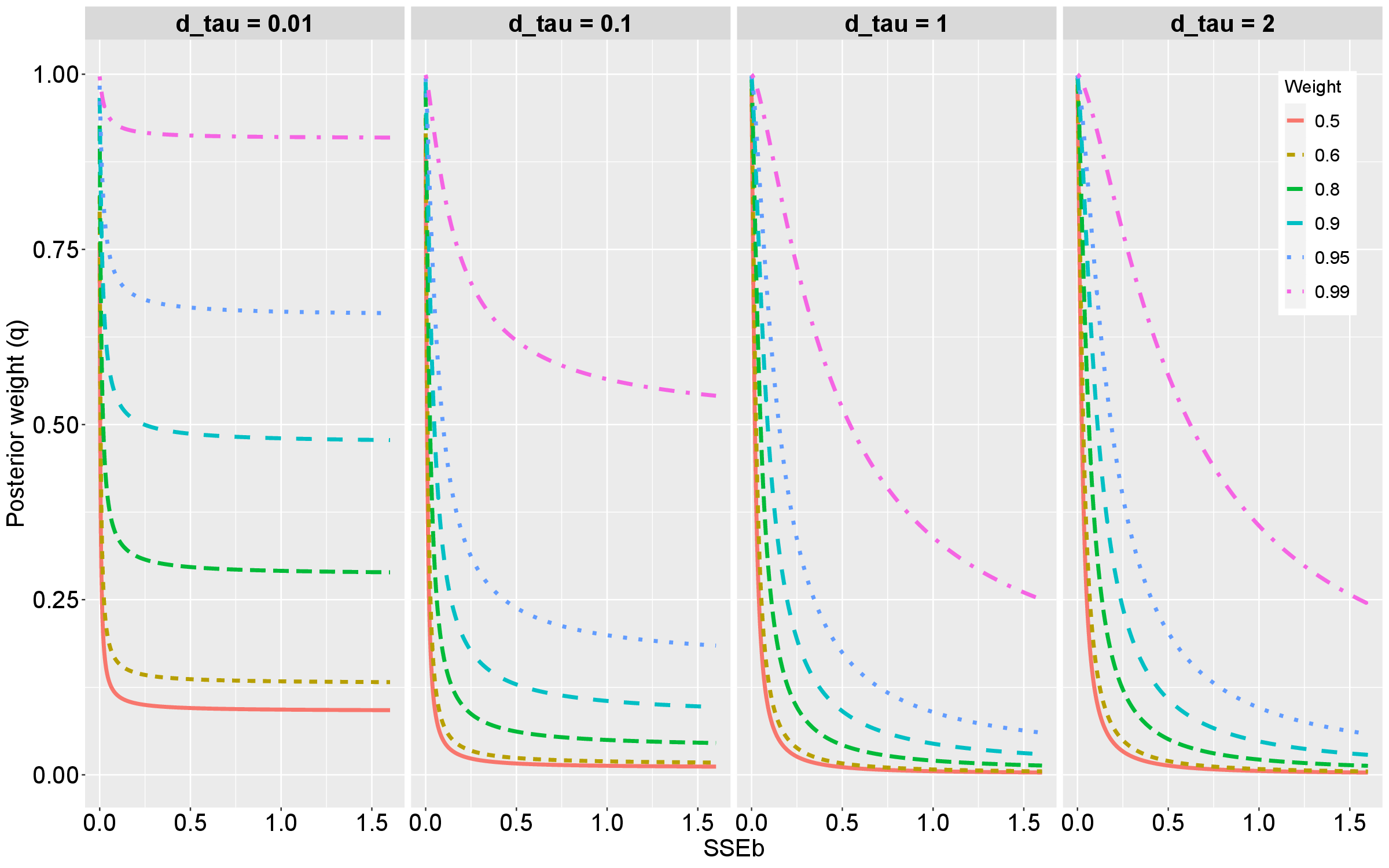}
	\caption{Plot of the of posterior weight  as a function of the sum of squared error (SSEb) of the log baseline hazards, for the inverse gamma mixture prior (\ref{eq:tau_mix}) for different values of $d_\tau$. The hyperparameters are; $a_\tau = c_\tau=1$, $b_\tau = 0.001$ and $d_\tau$ is either 0.01, 0.1, 1, or 2. The prior weight $p_0$ is 0.5 (red), 0.6 (orange), 0.8 (green), 0.9 (turquoise), 0.95 (blue) and 0.99 (maroon).}
	\label{Fig:qpost}
\end{figure}

\begin{figure}[H]
	\centering
	\includegraphics[width=14cm]{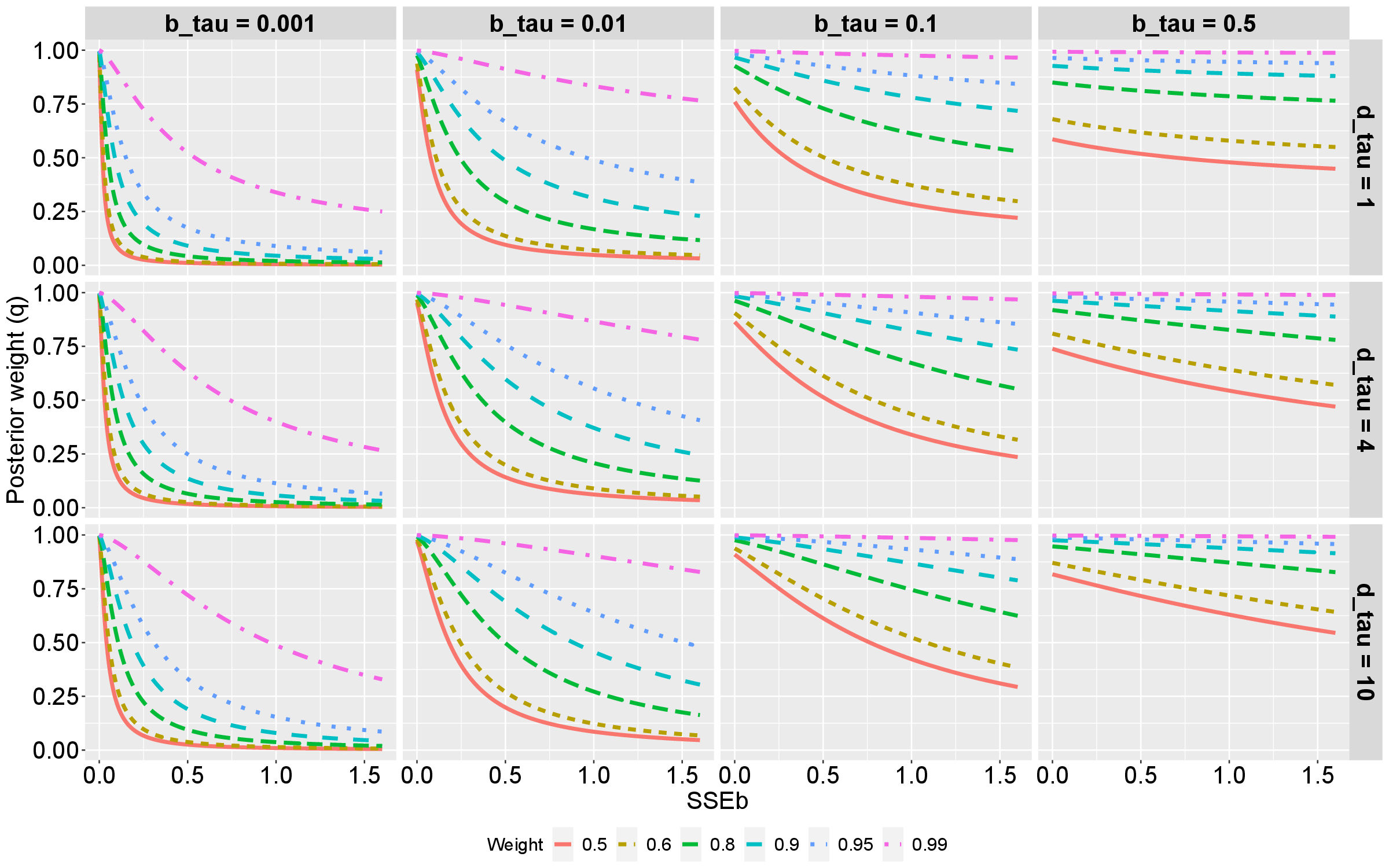}
	\caption{ Plot of the of posterior weight as a function of the sum of squared error (SSEb) of the log baseline hazards, for hyperparameters of $b_\tau$ of 0.001, 0.01, 1 and $d_\tau$ of 1, 4 10. The prior weight $p_0$ is 0.5 (red), 0.6 (orange), 0.8 (green), 0.9 (turquoise), 0.95 (blue) and 0.99 (maroon).}
	\label{Fig:qpost_db}
\end{figure}

To select suitable hyperparameters, we examine the dependence of $q_j$ on 
SSEb in terms of various hyperparameter combinations via expression 
(\ref{eq:qMSE}) and Figures \ref{Fig:qpost} and \ref{Fig:qpost_db}. The 
plots of $q_j(\text{SSEb}|\cdot)$ are the borrowing profiles, as they define how quickly historical
information is discounted for discrepancies between the log baseline hazards. We describe how the choice of hyperparameters affects the support and shape of the profile. 

The choice of $b_\tau$ is critical in ensuring that the support of $q_j(\text{SSEb}|\cdot)$ is close to (0,1). If $b_\tau$ is not sufficiently small, the support of $q_j(\text{SSEb}|\cdot)$ can be truncated depending on the choice of prior weight $p_0$. Conveniently, a choice of $b_\tau =0.001$ to encourage borrowing and $d_\tau \geq 1$ has the additional benefit of ensuring a full support $q_j(\text{SSEb}|\cdot)$ regardless of the prior weight. 


From a dynamic borrowing perspective, the ideal shape of the borrowing profile would be a step function of 1 for small tolerable differences in the SSEb and 0 otherwise.  Although this cannot be achieved, if we fix $d_\tau$ to a value of 1 or greater, setting $b_\tau$ to 0.001 and $p_0 < 0.95$ forces an appealing ``L" shape over an approximate (0,1) support for $q_j(\text{SSEb}|\cdot)$. The choice of $p_0$ then defines how quickly historical information is discounted for discrepancies between the log baseline hazards, through the reduction in the posterior weight. Increasing the prior weight $p_0$ and $d_\tau$, with $b_\tau$ fixed, increases the posterior weight for the same SSEb values.



At $q_j(\text{SSEb}|\cdot) < 0.5$, the smear mixture element is more likely, thus we consider $q_j(\text{SSEb}|\cdot)=0.5$ as a tipping-point. The mixture weight can then be defined in terms of the limit of tolerable differences between the log baseline hazards $\xi_j$. 
\begin{equation}\label{eq:tip_p}
p_{0j} = \left( 1 +  \frac{1}{d_\tau}  \left( \frac{\xi_j^2 + 2  }{\xi_j^2 + 2  d_\tau } \right)^{-\frac{3}{2}}
\right)^{-1}.
\end{equation}
This expression returns the same borrowing profile for $\xi_j$, by adjusting the prior weight for our choice of $d_\tau$. Larger values for $d_\tau$, for the same $\xi_j$, lead to a smaller prior weight $p_0$. 

Finally, we describe a principled approach to choosing the dynamic borrowing hyperparameters and interpret the prior in terms of an approximate sample size. To understand the scale of the baseline hazard function, the model is applied to the historical data set, and the smoothed baseline hazard plotted.  To encourage borrowing for tolerated differences, guarantee an appropriate lower bound for the posterior weight and ensure a quick reduction in borrowing for agreeable differences,  $b_\tau$ is set equal to 0.001 and $d_\tau \geq 1$. Its important that $d_\tau$ ensures the ``smear" in the prior is sufficient, to allow for the assumption of non-exchangbility, given the scale of log hazard. Once $d_\tau$ has been selected, a limit for tolerable differences between the log baseline hazards $\xi_j$ is defined and $p_{0j}$ calculated from (\ref{eq:tip_p}).   

For example, a 0.2 limit of tolerable difference between the log baseline hazards $(\xi_j = 0.2$) and a $d_\tau =1$, the commensurate prior (\ref{eq:tau_mix}) mixture weight is $p_0 = 0.75$. The prior
\begin{equation}
    \tau_j \sim (0.75)\text{IGamma}(a_\tau = 1, b_\tau=0.001) + (0.25) \text{IGamma}(c_\tau = 1, d_\tau=1)
\end{equation}
can be interpreted in terms of an approximate a priori ESS of $0.75n_0$ historical control patients. The posterior mixture weight $q_j(\text{SEEb}|\cdot)$ which defines the borrowing profile, with respect to this prior choice, is in the left panel of Figure 3 (Supplementary Material).  

The mixture prior on the variance of the log baseline hazards across the intervals (\ref{eq:tau_all}),  is explained in the same way in the Supplementary Material. The difference in the posterior update is that both the shape parameter and the scale parameter will be larger. The effects will tend to cancel each other out, so the interpretation of $q(\text{SSEb}|\cdot)$, rather than $q_j(\text{SSEb}|\cdot)$, is the same. The limit of tolerable difference $\xi_j$ in (\ref{eq:tip_p})  will be replaced by the limit of the sum of squared error. This will be harder to define, so the user may wish to plot various borrowing profiles for different choices of $p_0$ to decide on a suitable value.

\subsection{Posterior sampling}
The prior (\ref{eq:sp_prior}) for the number of split points $J$ in the piecewise exponential likelihood (\ref{eq:conc_like}) complicates the sampling of the joint posterior distribution. Since $J$ is a random variable, the dimension of the model parameter changes for each value in the support, standard MCMC software is not suitable. 

For fixed $J$ the unknown parameters in the joint model are 
\begin{equation}
    \bm{\theta}(J) =(\bm{\beta}, \bm{\lambda}, \bm{\beta}_0,  \bm{\lambda}_0, \mathbf{s}, \bm{\tau}, \sigma^2_\lambda, \mu
    ) . 
\end{equation}

To perform posterior estimation and inference, we use a Gibbs sampling algorithm to generate samples from the full posterior distribution. In the resulting MCMC scheme, there are a total of 10 moves. For fixed $J$, the components of $\bm{\theta}(J)$ are updated by either exploiting conjugacies in the full conditionals or via Metropolis-Hastings steps. Updating $J$ requires a change in the dimension of the parameter space; a RJMCMC algorithm \citep{Green1995} was developed and implemented. A detailed description of the complete algorithm, together with all necessary full conditional posterior distributions, is provided in the Supplemental material (Section 3). The algorithm has been implemented in the BayesFBHborrow package for R (R Development Core Team, 2012), available from the Comprehensive R Archive Network (CRAN, http://cran.r-project.org).

\subsection{ Summary of flexible baseline hazard model}
To summarise the flexible Bayesian model, we obtain the estimate of the historical baseline hazard $\lambda_0(t)$, and $\lambda(t)$ with the same approach, by an ensemble average
\begin{equation*}
    \mathbb{E}[\lambda_0(t)] = \frac{1}{m} \sum_{\nu=1}^{m} \lambda_{0j}(\nu),
\end{equation*}
where, $_j(\nu)$ in $\lambda_{0j}(\nu)$, is the interval containing $t$ in the MCMC iteration $\nu$ and $m$ is the number of iterations. The smoothness of this function is controlled by $\phi$ and $c_\lambda$. These are linked by a commensurate prior, where the variance parameter of $\tau$ or $\tau_j$ controls the borrowing. The overall variability of the historical baseline hazard is defined by $\sigma^2$. The log hazard ratios for the covariates $\vec{x}_{0i}$ and $\vec{x}_{i}$ are $\bm{\beta}_0$ and $\bm{\beta}_0$ respectively.

\subsection{Comparison approaches}
\label{Sec:Comp_app}

We compare our method in a setting where we are interested in borrowing the baseline hazard from a historical control arm for a model where we wish to make an inference on the marginal treatment effect. We make a comparison with two other Bayesian approaches which use the PEM with fixed time intervals, a simple informed prior (used within a mixture or by itself) and a hierarchical model. The mixture prior is motivated by \cite{Roychoudhury2020}, while the hierarchical prior is the parameterisation adopted in the FACTS software, commonly used within the pharmaceutical industry. 

A conjugate update is used to derive the informed prior. An appropriate estimate of the average hazard of death per unit time over each interval is the observed number of deaths in that interval,
divided by the average time survived in that interval. This latter quantity is
the average number of persons at risk in the interval, multiplied by the length
of the interval. Let the number of deaths in the $j$th time interval be $d_j$, $j =
1, 2, . . . ,m$, and suppose that $n'_j$ is the average number of individuals at risk
of death in that interval assuming that the censoring process is such that the
censored survival times occur uniformly throughout the $j$th interval 
\begin{equation*}
    n'_j = n_j - c_j/2
\end{equation*}
where $c_j$ is the number of censored individuals within the interval. Assuming that
the death rate is constant during the jth interval, the average time survived
in that interval is $(n'_j-d_j/2)\xi_j$ where $\xi_j$ is the length of the $j$th interval the hazard is estimated by
\begin{equation*}
    h^*(t) = \frac{d_j}{(n'_j-d_j/2)\xi_j}.
\end{equation*}

This leads to the prior 
\begin{equation*}
    \lambda_j \sim \text{Gamma}\left( n_jw , \frac{n_jw}{h^*(t)}\right)
\end{equation*}
where $w < 1$ weights the information, with smaller values increasing the dispersion around the estimated baseline hazard.

We also include a ``robustified" mixture prior to add a ``dynamic update". The informed prior is combined with a vague prior in the form of a mixture with an associated mixture weight.

The second comparator is a Bayesian hierarchical model (referred to as hierarchy in the results) where the hazard for individual $i$, in the $j$th interval for the $h$th trial is
in the form of 
\begin{equation}
\begin{array}{l}
    \lambda_{ijh} = \lambda_j  \exp\left\{\gamma_h + x_i \beta\right\} \nonumber\\
    \lambda_j \sim \text{Gamma}(a_\lambda , b_\lambda) \\
    \gamma_h \sim \mathcal{N}(\mu_\gamma, \tau^2) \\
    \mu_\gamma \sim \mathcal{N}(0, t^2_\gamma) \\
    \tau^2 \sim \text{IGamma}(a_\tau,b_\tau)\\
    \beta \propto 1 \nonumber
\end{array}
\end{equation}
where $\gamma_h$ is a frailty term for the study and $\lambda_j$ is the baseline hazard indexed by $h=0,1$ (historical and current) for fixed intervals $j = 1,...,J+1$. We use $\text{Gamma}$ for the gamma distribution with shape and rate hyperparameters $(a_\lambda, b_\lambda)$, $\text{IGamma}$ for the inverse gamma distribution with shape and scale hyperparameters $(a_\tau, b_\tau)$ and $\mathcal{N}$ for a Gaussian distribution with mean and variance hyperparameters.  

A conditionally conjugate Gamma prior is placed on each piecewise baseline hazard $\lambda_j$. The $\gamma_h$  terms are the relative study-level effects (random effect) on the hazard for the segment $s$. A hierarchical model is posited across these study-level effects, with $\gamma_h$ drawn from a normal distribution with hyperparameters $\mu_\gamma$ and $\tau^2$. 

Here, $\tau^2$ controls the overall borrowing in a similar way to (\ref{eq:tau_all}). The prior standard deviation for $\mu$ is set equal to the largest log hazard ratio of event rates for historical studies. The expectation for $\tau^2$ is set to the same value as the prior standard deviation for $\mu$. The updates and sampler are described in the Supplementary Material and the code is available on request. 

Although we apply the approach in the context of a single historical dataset, we appreciate that this model also extends to multiple historical datasets. However, the hierarchical structure is most effective in preserving the primacy of the current trial in the presence of a single historical dataset and becomes less ``dynamic" as we add more historical trials to the model. We can see this from the conditional conjugate update of $\tau^2$ (all model updates are in the Supplementary Material Section 5.2)
\begin{equation}
    \tau^2|\cdot \sim \text{IG}\left(a_\tau + 1, b_\tau+ \sum_h \frac{(\gamma_h-\mu_\gamma)^2}{2} \right)
\end{equation}
where $h$ indexes the current or historical trials. In the case of one historical trial, it is the discrepancy between the current and historical trial which determines $\tau^2$ and the discounting of the borrowing. As we add more historical trials, the primacy of the current trial diminishes, as the contribution of the current trial in the posterior update is overwhelmed by the historical data.

\section{Simulation study}
A simulation study is conducted to understand the impact of the smoothing hyperparameters for estimating the historical baseline hazard from our prior structure, asses how each proposed prior on $\tau$ in our Bayesian flexible baseline hazard model (FBHM) impacts the borrowing from the historical control and to compare the performance against alternative approaches.

Under a $90\%$ power and two-sided type I error of $5\%$,  a total of 265 deaths in the trial is required (based on a log-rank test) with $10\%$ censoring to detect a reduction in baseline hazard by $40\%$ for balanced treatment to control allocation. Rather than a single trial, we simulate a current trial with 150 patients in the treatment group and 100 in the control group. The deficit of patients within the control group is  made up from a historical study which contains 100 patients. A total of 500 trials are simulated for three different shapes of the baseline hazard under four different scenarios; 
\begin{itemize}
    \item A. Null case where the baseline hazard is the same for historical and current with no treatment effect. 
    \item B. Baseline hazard is the same for historical  and current with treatment effect on the baseline hazard of either $\beta = \exp(-0.5)$ (scenario B1) and $\beta = \exp(-0.275)$ (scenario B2).
    \item C. Prior-data conflict. Baseline hazard of the historical trial is lower than hazard of the current trial with no treatment effect, leading to a potential false conclusion of a treatment effect if too much borrowing occurs.  
    \item D. Prior-data conflict. Baseline hazard of the current data is greater than the historical  trial, in the presence of a treatment effect of $\beta = \exp(-0.275)$, leading to a loss of power.  
\end{itemize}

To illustrate the flexibility of the model to borrow across a variety of baseline hazards, we simulate data from a Weibull distribution with a monotonically increasing hazard (shape $(\kappa) = 1.5$, scale $(\nu) = 0.4$) and a Weibull mixture $0.5\text{Weibull}(\kappa_1 =0.6, \nu_1=1.2) + 0.5\text{Weibull}(\kappa_2 = 2.5, \nu_2 = 0.3)$ with a non-monotonic hazard. We adjust the parameters according to the scenario, the corresponding baseline hazards are in Figure 4 (Supplementary Material). In scenario C, the shape parameter ($\kappa_1$ in the mixture) of the Weibull distribution for the historical data simulation is reduced by $15\%$. In Scenario D, the same parameter is increased by $15\%$ for the current data. For all datasets, censoring is predominantly from loss to follow-up.



To understand how the prior structure estimates the baseline hazard and make good choices in setting the hyperparameters, the first 100 historical datasets from the null scenario of the Weibull and mixture Weibull simulated dataset are applied to the model without the borrowing structure. The smoothing parameter is fixed at values between 0 and 1 for three different values of $\phi$ (the parameter in the truncated Poisson, which is approximately the mean number of split points), 3, 7 and 10. The mean squared error (MSE), standard deviation, and quartile coefficient of dispersion are calculated from the posterior baseline hazard to evaluate the model. 

Performance of the borrowing prior structure is evaluated using power and type I error, average bias of the treatment difference, empirical coverage probabilities and mean-squared-error (MSE) of the log current baseline hazard. A treatment effect is declared if the upper bound of a two-sided $95\%$ highest posterior density credible interval is below 0. We focus on the borrowing aspects of the model, so only include a treatment as a covariate within the simulated dataset. 

We use the following hyperparameters for our Bayesian flexible baseline hazard model (FBHM) in the Weibull simulation; $a_\tau = c_\tau = 1, b_\tau = 0.001$ and $d_\tau = 1$. We set the mixture weight for both the FBHM and the informed prior to $p_0 = 0.5$. This corresponds to a strict ``L" shape borrowing profile (Figure \ref{Fig:qpost_db}) and a 0.134 difference between the log baseline hazards ($\xi_j$ for the $\tau^{(\text{mix})}$ prior) at the tipping point.   

In the mixture of Weibull distributions $d_\tau$ is set to 10 to reflect the possibility of larger baseline hazard values at earlier time points. This increases the difference between the log baseline hazards ($\xi_j$)  at the tipping point to 0.447 while maintaining the ``L" shape borrowing profile. The approximate prior mean $(\phi)$ and maximum number of split points are set to 3 and 5 respectively. The multivariate Gaussian hyperparameters are fixed at $a_\sigma = b_\sigma = 1$, with $c_\lambda$ of 0.7 and 0.3 for the Weibull and mixture distribution, respectively. 

We compare our FBHM with two Bayesian PEM models that dynamically incorporate historical data (explained in Section \ref{Sec:Comp_app}) and one Bayesian approach which does not. For these comparison approaches, we use a piecewise exponential likelihood with fixed time intervals, the intervals are set at the $100(k/K)^{\text{th}}$ percentile of observed failure times, with $K = 3, 4, 5$. 

We perform eleven thousand five hundred Monte Carlo iterations, with the first 1500 discarded as burn-in. The baseline mean squared error for the simulations is calculated by 
\begin{equation}
    \text{MSE}(\lambda) \approx \sum_{t \in t^*} \frac{(\lambda(t) - \hat{\lambda}_1(t))^2}{2000}
\end{equation}
where $t^*$ is an equally spaced grid of 2000 time points from close to zero to $\text{max}(\vec{y})$ and $j$ depends on the baseline line hazard.  In the mixture Weibull simulation, we adjust the grid from 0 so that the very large spike of the baseline hazard at the start of time does not affect the MSE estimation.



\subsection{Results from current study only}

\begin{figure}[H]
	\centering
	\includegraphics[width=14cm]{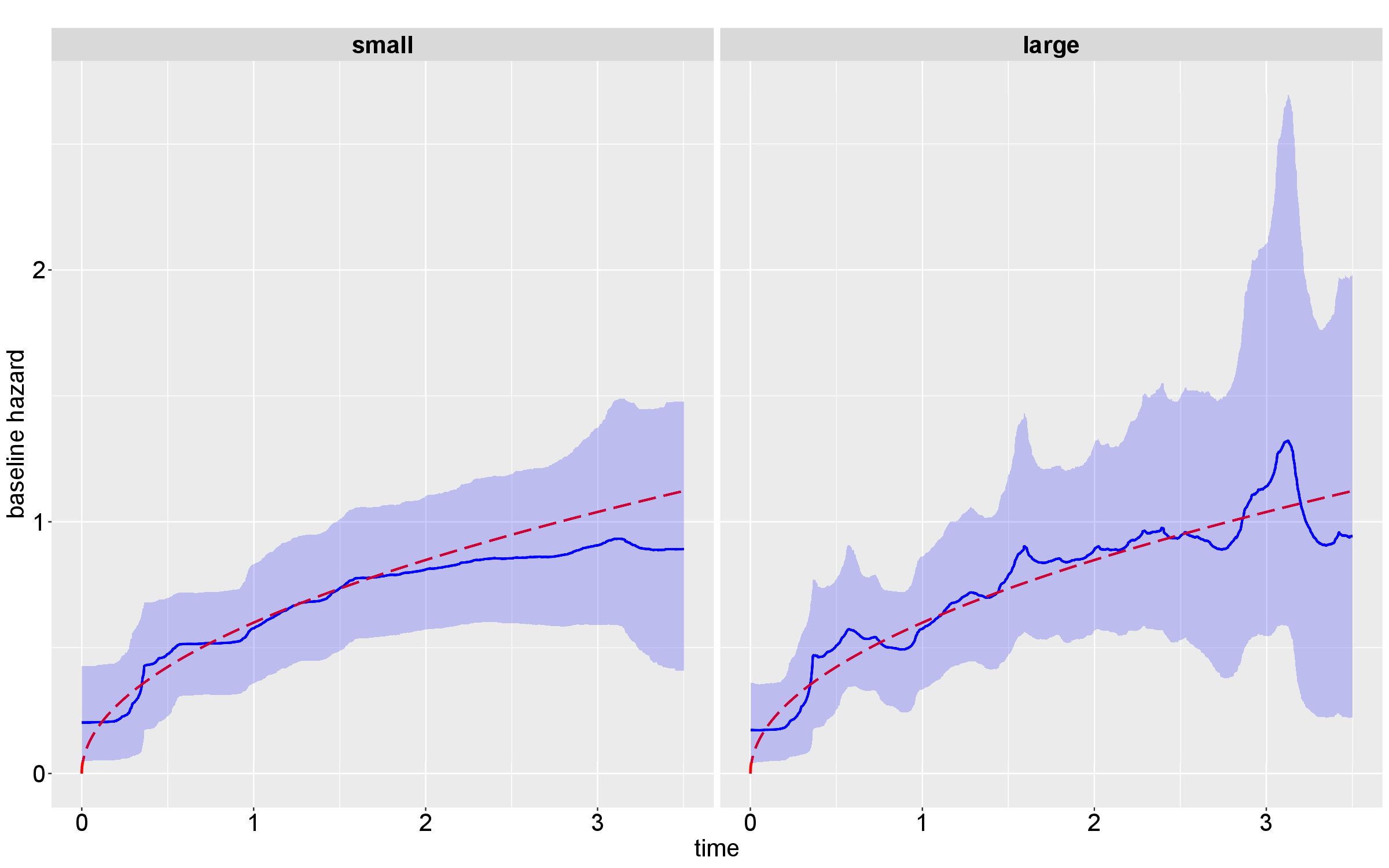}
	\caption{Simulation study current data only: Plot of the smoothed posterior baseline hazard (blue), 90$\%$ credible intervals (light blue) and the true baseline hazard (dashed red) for $c_\lambda = 0.6$ for $(\phi=3$, $J_{\text{max}}=5$) labelled small and  ($\phi=10$, $J_{\text{max}}=20$) right labelled large for one simulated dataset from the Weibull distribution.}
	\label{Fig:SmoothEx}
\end{figure}

Figure \ref{Fig:SmoothEx} shows the posterior baseline hazard function from the ensemble average for the simulated data using smoothing priors, showing a good approximation of the true hazard. As expected, the $90\%$ credible interval increases as the later time points. There is a trade-off between accuracy and uncertainty of the baseline hazards within the model; at the earlier time points, where there is plenty of information, accuracy of the estimated baseline hazard is increased by more split points, where as at later time points the uncertainty of the baseline hazards is reduced at later time points by fewer split points. The baseline hazard estimate is more accurate and less variable with a small prior mean of split points. The lowest MSE and standard deviation for the posterior baseline hazard across all time points are from the model with $\phi$ of 3 (and a maximum number of split points of 5) regardless of the underlying shape of the baseline hazard. Although this reduces the flexibility of the smoothed posterior baseline hazard, the wider time intervals within the PEM ensure that there is more data to estimate the baseline hazard at later time to events, where there are fewer events, reducing the overall variability of the baseline hazard estimate.  The restriction on the split points to a small number appears to produce a more regular shape and stops the model from overfitting the posterior baseline hazard to the data, as seen in Figure \ref{Fig:SmoothEx}.  

The total number of time points becomes more important when the censoring is primarily random, where the sparsity of events at later time points is much larger. Allowing a large number of split points leads to intervals near $s_{\text{max}}$ where no event has occured, inducing greater variability in the posterior baseline hazard and increased MSE.   

\begin{figure}[H]
	\centering
	\includegraphics[width=14cm]{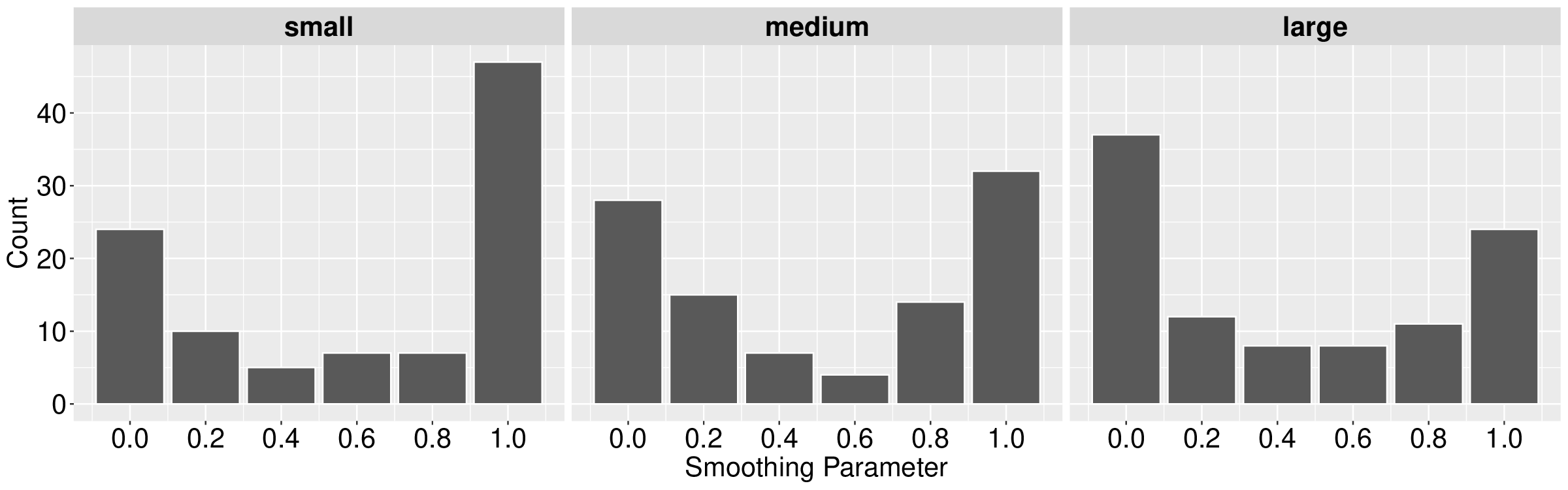}
	\caption{Simulation study current data only: Bar charts displaying the optimal choice of $c_\lambda$, in achieving the smallest MSE for the estimated baseline hazard  for 100 simulated datasets from the Weibull distribution for the three different $(\phi, J_\text{max})$ small (3, 5), medium (5, 10) and large (10, 20). As the number of allowable split points gets larger and the model becomes more flexible the smoothing parameter is less effective and improving the model fit. }
	\label{Fig:wei_nb_barcharts}
\end{figure}

Both $c_\lambda$ and $\phi$ control the smoothing of the baseline hazard. For a small prior mean of split points and a monotonic baseline hazard, $c_\lambda$ is important. The average MSE (across the simulations) is smallest for $c_\lambda = 0.6$ and the modal value of $c_\lambda$ that achieves the lowest MSE is 1 (Figure  \ref{Fig:wei_nb_barcharts}). A larger $\phi$, encourages more split points, reducing the impact of $c_\lambda$. This can be seen for $\phi$ of 10, where the lowest average MSE is smallest for a $c_\lambda$ of 0. For the non-monotonic hazard function, the smoothing is not necessary.

An underlying pattern of the results is that for most samples, the smoothing parameter $c_\lambda$ either increasingly reduces or increases the MSE. Given that the optimal setting for the split points is $\phi$ of 3, the optimal choice of $c_\lambda$ can be made with an understanding of the shape of the baseline hazard. Typically, a fairly regular shape is expected, so a large value of $c_\lambda$ would be optimal. 

\subsection{Results with borrowing from historical data}

A key aspect of the results is that regardless of the shape of the baseline hazard, the FBHM leads to a much more accurate estimate of the baseline hazard compared with the hierarchical prior, allowing the model to exploit the flexible parameterisation. This leads to increased power when borrowing is appropriate and to a reduction in the bias of the estimated treatment effect. 

Our FBHM has larger power than the hierarchical and mixture prior with fixed time points in scenario B (Tables \ref{Tab:WeiBD_fbhm} - \ref{Tab:WeiB2} and Tables \ref{Tab:WeiMixBD_fbhm} - \ref{Tab:WeiMixB2}), where the baseline hazard for the historical and control is identical for both treatment effects (B1 and B2). The informed prior achieves the highest power and the flat prior the lowest in the Weibull simulation; this is to be expected when the samples from the historical control are from the exact distribution of the current trial, as the best approach is simply to use all the data in the same model (Tables \ref{Tab:WeiA}, \ref{Tab:WeiMixA}). For the mixture simulation, the power of the informed approach is below the two comparator borrowing approaches as it is difficult to model the non-monotonic baseline hazard with a step-function. Our method (FBHM) outperforms all the other approaches regardless of the prior.

The difference in power for three borrowing priors in our FBHM highlights the importance of the choice of weight $p_0$ and vague component $d_\tau$ in the mixture prior. This is addressed by the principled approach to choosing their values outlined in Section \ref{sec:pa_hyper}. The choice of $p_0$=0.5 and $d_\tau=1$ defines a conservative limit for the tolerable differences between the log baseline hazards ($\xi_j$) and leads to a reduction in power for the Weibull simulation. In the mixture simulation, the choice of $d_\tau = 10$ and $p_0=0.5$, defines a large limit for $\xi_j$ and a flatter borrowing profile, increasing both the power and type I error. This profile can also be achieved with a smaller weight while keeping $d_\tau$ at 1. 

The FBHM has a substantially lower type I error compared to the two other dynamic borrowing prior structures in scenario C (Tables \ref{Tab:WeiAC_fbhm}, \ref{Tab:WeiC} and Tables \ref{Tab:WeiMixAC_fbhm}, \ref{Tab:WeiMixC}), where the baseline hazard of the historical trial is lower than the baseline hazard of the current trial without a treatment effect. In the Weibull simulation the hierarchical prior has as much as a $11.4\%$ type I error compared to $4\%$ for the FBHM with the mix prior. This lower type I error does not come with a lower power for the FBHM in scenario B or scenario D (Tables \ref{Tab:WeiBD_fbhm}, \ref{Tab:WeiD} and Tables \ref{Tab:WeiMixBD_fbhm}, \ref{Tab:WeiMixD}), where the difference in baseline hazards decreases the effect of borrowing in the presence of a treatment effect. The hierarchical approach marginally outperforms the informed mixture prior and flat prior in scenario D, with a lower mean bias of the treatment effect and 0.001 increase in power. The FBHM is still able to maintain a power close to the optimal power and has a $50\%$ lower bias. This pattern is also observed in the case of the mixture Weibull simulation. However, the change in the true baseline hazard is less pronounced, thus the power reductions for the informed mixture prior and hierarchical prior are less.   

In the case of the Weibull simulation studies, the informed mixture prior outperforms the hierarchical prior for the low treatment effect and matching baseline hazards (scenario B1) in the Weibull simulation, but still maintains a lower type I error for a mismatch in baseline hazards with no treatment effect (scenario C). This pattern is not replicated in the mixture simulation where the hierarchical prior has a slightly large power for scenario B1, but a larger type I error for smaller split points. 

The hierarchical prior is less susceptible to differences in results from the choice of split plots, but there is no discernible pattern in terms of the optimal number of splits, this depends on the unknown shape of the underlying baseline hazard. There can be quite large differences in results with the informed mixture prior, there is a $4.6\%$ differential in type I error between the 3 and 5 splits in the Weibull simulation. This is not as pronounced for the hierarchical prior.   

Clearly, estimating the underlying baseline hazards accurately is an important factor to ensure appropriate dynamical borrowing from the control. This is made clear in scenario B2, where the non-monotonic baseline hazard for the current and historical match for a moderate treatment effect. The FBHM increases power beyond simply combining the likelihoods because the borrowing is based on a much more accurate approximation of the true shape of the underlying baseline hazard.  However, optimising this within the FBHM does not lead to the best performance characteristics. The results imply that the mixture of inverse gamma prior on $\tau$ (\ref{eq:tau_mix}) or $\tau_j$ (\ref{eq:tau_all}) improves the fit of the model and generally achieves better overall borrowing characteristics. The mixture prior parametersiation across all the intervals (\ref{eq:tau_all}) leads to the best model fit regardless of the shape of the underlying baseline hazard. However, the mixture prior for each interval (\ref{eq:tau_mix}) always achieves a lower type I error, regardless of the underlying shape of the baseline hazard. The choice of the mixture weight and the vague component is important in determining the level of borrowing.  

The box plots in Figures \ref{Fig:bp_sc24} and \ref{Fig:ds_bp24} (and Figures 6, 7 in the Supplementary Material) show that the estimation of the underlying baseline hazard is much more accurate with the FBHM compared to the comparison approaches. The hierarchical model has a large MSE which is very susceptible to the choice of split points. In both simulation studies, as the number of split points increases, the MSE for the hierarchical prior gets larger.

\section{Application}
We demonstrate our method in a clinical trial setting with two phase 3 trials in the same target population. As Bayesian borrowing is typically implemented in phase II trials, we mimic this setting by subsetting the trial data. The placebo control in the target trial is supplemented by borrowing external placebo patients from the historical trial within our FBHM.  

The placebo-controlled target trial is for the efficacy of benralizumab (SIROCCO) \citep{Bleecker2016}, a monoclonal antibody for the treatment of severe eosinophilic asthma. Eosinophilia is present in approximately $50\%$ of asthma patients and is associated with a worsening of asthma severity and decreased lung function, with an increased risk of exacerbation. The trial was conducted at 374 sites in 17 countries with recruited patients (aged 12 to 75 years) with a physician-based diagnosis of asthma for at least 1 year and at least two exacerbations while on oral corticosteroids in the previous year. In our setting, adult patients (18+) are randomly assigned (1:1) to benralizumab 30 mg every 8 weeks or placebo for 48 weeks as an add on to their standard treatment. Patients' recruitment was stratified 2:1 according to blood eosinophil counts of at least 300 cells per $\mu_\text{L}$ and less than 300 cells per $\mu_\text{L}$. 

The primary endpoint was the annual exacerbation rate ratio versus placebo, but we focus on a key secondary endpoint of time-to-first exacerbation. The time-to-first asthma exacerbation for patients who did not experience an asthma exacerbation during the treatment period was censored on the date of their last visit for the 48-week double-blind treatment period or at the time point after which an exacerbation could not be assessed (for lost-to-follow-up patients). 

External adult patients are from the placebo-controlled arm of the trial evaluating benralizumab (CALIMA) \cite{FitzGerald2016} as an add-on for patients with severe uncontrolled eosinophilic asthma. Patients are in the same age group to the target population, and were again stratified (2:1) by baseline blood eosinophil counts of 300 cells per $\mu_\text{L}$ or greater and less than 300 cells per $\mu_\text{L}$, respectively. The model assumption of exchangability on the difference in log hazards between the historical and current data sets seems plausible. To pool the information for the targeted estimand, patients are censored after their follow-up period exceeds the maximum time to follow-up of the current trial (50 weeks). 

The sample size calculation, assuming a two-sided type I error of $10\%$ and a power of $80\%$ to detect an increase in survival probability at 48 weeks of $40\%$, gave a requirement of 84 patients per arm (see Supplementary Material). Rather than ``recruit" the 84 control patients within the target study, the sample size of the control group is reduced to 42 patients and supplemented by borrowing from the placebo-controlled patients of the CALIMA study, stratified (2:1) as in the target study by baseline blood eosinophil counts. Time-to-first asthma exacerbation over 48 weeks is analysed in the intention-to-treat population with adjustment for treatment, region, exacerbations in the previous year, and oral corticosteroid use at the time of randomisation. These covariates are present in both current and historical studies and are included in the model. To avoid restricting the information in the historical data, care must be taken in the construction of the design matrix.  Whereas the treatment variable can be coded with respect to the reference group, all other discrete covariates must be coded to imply a sum-to-zero constraint on the regression parameters. As the proportions in the strata are the same across the studies, we do not need to reweight the design matrix to maintain the same interpretation of the baseline hazard.  

First, we run our FBHM model on the CALIMA (historical) data only, to assess the choice of smoothing parameters and appropriate tipping-point. The baseline hazard appears to be a fairly regular shape, so we control the smoothness of the estimated baseline hazard set to $\phi=3$, $J_\text{max}=5$ and $c_\lambda=0.6$ to achieve a balance between flexibility and uncertainty. The hyperparameters for variability of the historical log baseline hazard, $a_\sigma$ and $b_\sigma$, are set to 1.   

Two different models are explored which borrow to supplement the recruitment of patients in the control arm; (1) we wish to limit the effective sample size a priori to $n_0 = 42$, (2) we control the sensitivity of the borrowing model to tolerate differences between the log baseline hazards of less than 0.22. The borrowing is controlled via different choices for the prior weight $p_0$ in the ``mix" borrowing prior, with each generating 20,000 posterior samples after a burn in of 1,000. In the first model, we set $p_0 =0.5$, $b_\tau = 0.001$ and $d_\tau = 10$, for an ESS of $n_0 = 42$. This choice of hyperparameters defines the borrowing tipping point for the difference between the log baseline hazards $\xi_j$ as 0.2. In the second model, we define a borrowing tipping-point of $\xi_j = 0.33$, for a $p_0$ of 0.8. 

The variance proposals for $\bm{\beta}_0$ and $\bm{\beta}$ are tuned on short initial runs, so the acceptance rate for the treatment effect MCMC chain is approximately 40$\%$. Regardless of our choice of $p_0$, we are able to declare a treatment effect as the upper $95\%$ highest posterior density credible interval is below 0. The results, including the posterior mean, are in Tables 1 and 2 of the Supplementary Material. The smoothed posterior baseline hazard and survival function for time-to-first exacerbation, with associated $95\%$ credible intervals for $p_0 = 0.5$ are shown in Figure 8 in the Supplementary Material.  

We quantify the effect of borrowing by comparing the precision of the two treatment effects with and without borrowing. Borrowing the information from the historical trial improves the precision of the treatment estimate compared to a vague prior. For $p_0 =0.5$, this is equivalent to adding 80 patients to the treatment arm for the undersized trial.  We perform a counterfactual trial with no borrowing, where the full 84 patients (supplementing the same 42 control patients) are selected from the SIROCCO target trial.  A significant effect is also found, but the precision is 20\% higher than for the more liberal borrowing model (where $p_0 = 0.8$). This highlights the delicate balance between borrowing sufficiently to power the trial, and building in sufficient protection against prior-data conflict, when using historical information to supplement the control arm.

\section{Discussion}

If there is a high chance that the historical control is similar to the current control, using the information offers two clear advantages. Either we can reduce the recruitment on the current control and maintain the target power or boost the power of the trial to detect a treatment effect. In order to protect against possible differences between the two datasets such as drift over time, a Bayesian borrowing approach, which dynamically borrows information, will reduce the risk of inflation of type I error. By integrating over the uncertainty of the location and number of split points in our ensemble method and incorporating a smoothing prior, we remove any constraints on the shape of the baseline hazard in our joint semiparametric model. This leads to an improvement in the characteristics of the borrowing over current dynamic approaches which constrain the shape of the hazard function. The use of a mixture of inverse gamma distributions on the commensurability parameter $\tau$ for the FBHM allows the borrowing to be more ``robust'' to prior-data conflict. The choice of the hyperparameters is important in determining the sensitivity of borrowing to differences between the historical and current log baseline hazards, which in turn effects the amount of information is borrowed from the historical data. We show how these can be selected in a principled way using a tipping-point for tolerable differences and interpreted in terms of an ESS.    

Our simulation study suggests that an ideal choice for the model is to set a small number of possible split points (a combination of $\phi=3$ and $J_{\text{max}} = 5$ seems reasonable) and use $c_\lambda \in [0,1]$ according to the anticipated shape of the baseline hazard. Regular shapes with gradual changes require more smoothing, hence a value closer to 1 for $c_{\lambda}$, where as irregular shapes need less, so a value closer to 0. 

The joint hierarchical structure of the FBHM can be extended to multiple historical data sets by augmenting the model with the corresponding likelihood and adjusting the prior structure. The introduction of multiple datasets allows for more varied prior structures to capture the between-trial heterogeneity. With a similar structure to our commensurate prior (\ref{eq:tau_all}), \cite{Neuenschwander2016} proposed an extension to allow for non-exchangability. However, the effectiveness of the joint model to give primacy to current control patients diminishes as the number of historical data sets increases.  

Alternatively, a Dirichlet process mixture prior \citep{Escobar1995} could be used to account for departures from the exchangeability assumption by placing a prior on a distribution rather than a parameter, resulting in a discrete posterior distribution which creates a data-dependent clustering mechanism.  This approach has been adopted by \cite{Hupf2021} in the context of a binary end point and in a time-to-event setting \citep{Bi2023}. As the number of historical trials is usually small, this type of clustering can also be achieved using a RJMCMC approach.  

Our model requires IPD from the external control arm. This is obviously harder to obtain with one historical trial compared with group-level data and increasingly difficult if we wish to incorporate multiple trials. The computational aspect also increases when we add multiple trials. Depending on the amount of data available, alternative approaches such as variational inference should still allow for a flexible model whilst maintaining a workable computational speed.

\section{Plots}

\begin{figure}[H]
	\centering
	\includegraphics[width=14cm]{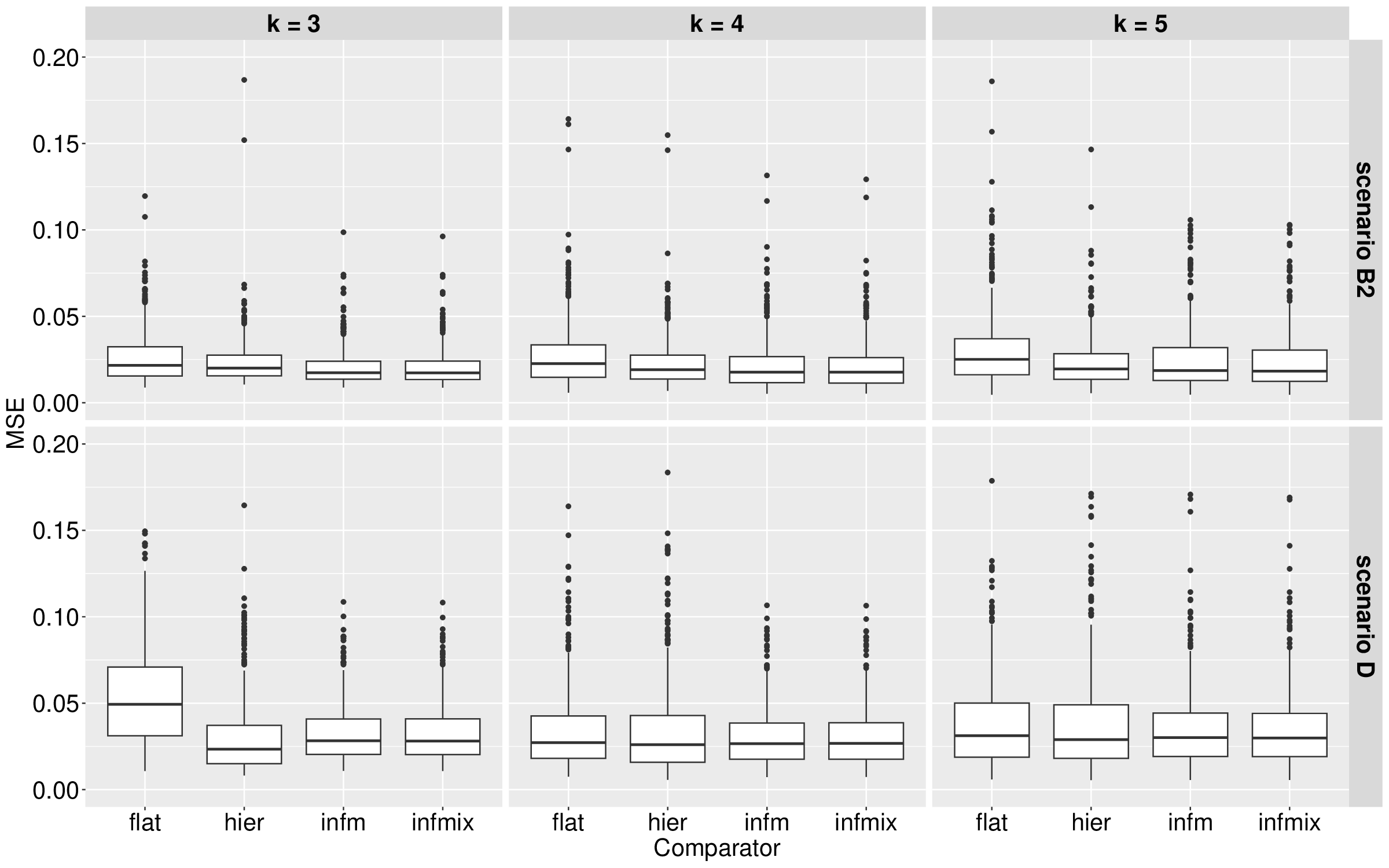}
	\caption{Comparison approaches for scenario B2 and D: Box plots of the MSE of the current baseline hazard for the comparison approaches  with fixed time intervals at the $100(k/K)^{\text{th}}$ percentile of observed failure times $K = 3,4,5$ for scenario B and D.}
	\label{Fig:bp_sc24}
\end{figure}

\begin{figure}[H]
	\centering
	\includegraphics[width=14cm]{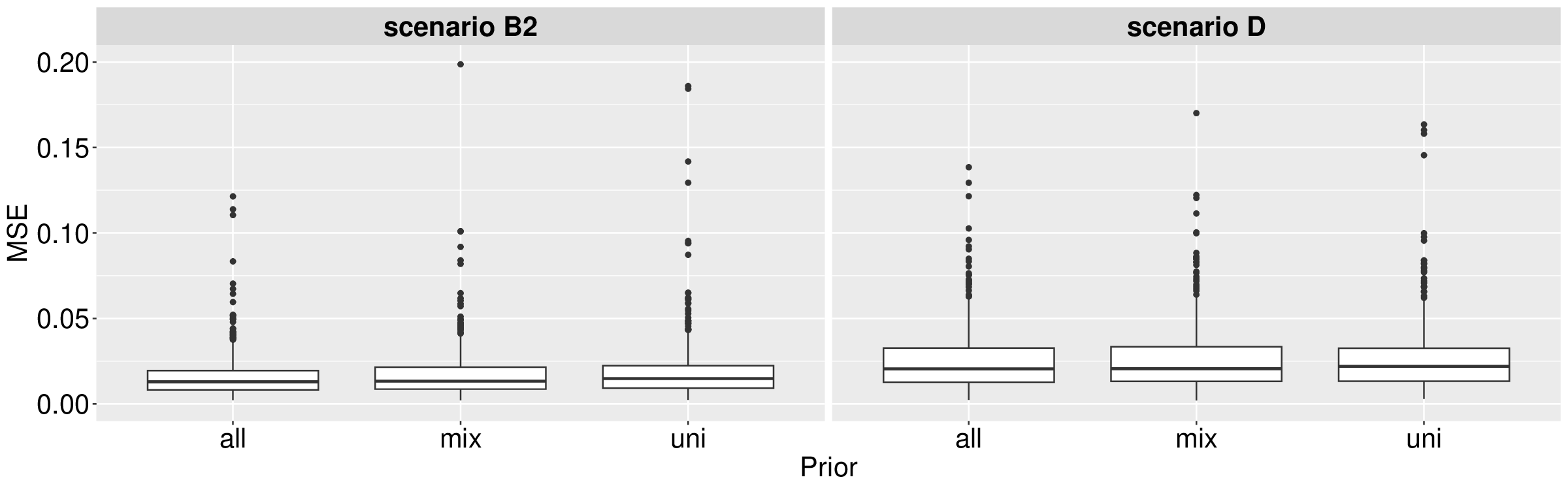}
	\caption{Flexible baseline hazard model for scenario A and C: Box plots of the MSE of the current baseline hazard for the FBHM with the 3 different prior structures all, mix and uni for scenarios B2 and D.}
	\label{Fig:ds_bp24}
\end{figure}

\printbibliography

\newpage
\section{Tables}
The tabled results from the 500 simulated trial datasets for scenarios A to D, where the baseline hazards are from either a Weibull or a Weibull mixture for all of the Bayesian models are below.  All tables contain the averaged difference between the estimated treatment effect and the true value (bias), the average standard deviation of the estimated posterior treatment effect (beta sd), the estimated current posterior baseline hazard (mse) and  average quartile coefficient of dispersion (qv) for each model. 

The tables either contain the results of the comparator Bayesian approaches or the FBHM for a particular scenario set. Scenario A and C tables have the type I error (type1) and empirical coverage  where as scenario B and D tables have the power (as the coverage is 1).  

\subsection{Weibull Simulation Results}
\begin{table}[H]
	\centering
    \begin{filecontents*}{ds_tab_sc1sc3.csv}
bias,beta sd,mse,qv,type1,coverage,scenario,prior
-0.0199,0.1212,0.0175,0.1261,0.014,0.986,A,all
-0.0179,0.1233,0.0196,0.1349,0.016,0.984,A,mix
-0.0477,0.1142,0.021,0.1318,0.018,0.982,A,uni
0.0432,0.1222,0.0155,0.1264,0.048,0.952,C,all
0.0369,0.1249,0.0179,0.135,0.04,0.96,C,mix
0.0349,0.1151,0.0193,0.1314,0.052,0.948,C,uni
\end{filecontents*}
\csvautotabular{ds_tab_sc1sc3.csv}
    \caption{Results from scenario A and scenario C for our Bayesian FBHM with the different priors on $\tau$ for the simulated data from the Weibull distribution.}
\label{Tab:WeiAC_fbhm}
\end{table}

\begin{table}[H]
	\centering
    \begin{filecontents*}{other_tab_sc1.csv}
bias,beta sd,mse,qv,type1,coverage,splits,approach
-0.0027,0.113,0.0484,0.0789,0.026,0.974,3,informed
-0.001,0.1107,0.0405,0.083,0.04,0.96,4,informed
0.0031,0.1087,0.0409,0.0872,0.044,0.956,5,informed
-0.0022,0.1165,0.0185,0.1039,0.028,0.972,3,infmix
-0.0027,0.1152,0.0162,0.1044,0.032,0.968,4,infmix
-0.0043,0.1141,0.018,0.1071,0.03,0.97,5,infmix
0.0067,0.137,0.0241,0.2365,0.022,0.978,3,flat
0.0069,0.1369,0.0229,0.2168,0.02,0.98,4,flat
0.0066,0.1369,0.0259,0.2066,0.02,0.98,5,flat
0.0028,0.1165,0.0204,0.0757,0.026,0.974,3,hierarcal
0.0039,0.116,0.0176,0.0851,0.026,0.974,4,hierarcal
0.005,0.1174,0.0187,0.094,0.024,0.976,5,hierarcal
\end{filecontents*}
\csvautotabular{other_tab_sc1.csv}
    \caption{Results from scenario A for the comparison approaches for the simulated data from the Weibull distribution.}
\label{Tab:WeiA}
\end{table}

\begin{table}[H]
	\centering
    \begin{filecontents*}{other_tab_sc3.csv}
bias,beta sd,mse,qv,type1,coverage,splits,approach
0.0794,0.1124,0.0641,0.0782,0.122,0.878,3,informed
0.0937,0.1097,0.059,0.0818,0.168,0.832,4,informed
0.1055,0.1078,0.0605,0.0858,0.21,0.79,5,informed
0.0662,0.1183,0.013,0.1103,0.076,0.924,3,infmix
0.0711,0.1169,0.0148,0.1099,0.102,0.898,4,infmix
0.0766,0.1159,0.0159,0.112,0.118,0.882,5,infmix
0.0126,0.1368,0.022,0.2366,0.03,0.97,3,flat
0.0128,0.1371,0.0261,0.217,0.034,0.966,4,flat
0.0126,0.137,0.0286,0.2067,0.032,0.968,5,flat
0.0849,0.1168,0.0129,0.0761,0.108,0.892,3,hierarcal
0.0879,0.1171,0.0127,0.0855,0.112,0.888,4,hierarcal
0.0897,0.1171,0.0135,0.0942,0.122,0.878,5,hierarcal
\end{filecontents*}
\csvautotabular{other_tab_sc3.csv}
    \caption{Results from scenario C for the comparison approaches for the simulated data from the Weibull distribution.}
\label{Tab:WeiC}
\end{table}

\begin{table}[H]
	\centering
    \begin{filecontents*}{ds_tab_sc2sc4sc5.csv}
bias,beta sd,mse,qv,power,scenario,prior
0.2039,0.1239,0.0163,0.1263,0.66,B1,all
0.2036,0.1262,0.017,0.1346,0.652,B1,mix
0.1746,0.1171,0.0183,0.1311,0.796,B1,uni
-0.0244,0.1274,0.0168,0.1269,0.988,B2,all
-0.0248,0.1298,0.0175,0.1358,0.988,B2,mix
-0.0549,0.1207,0.0194,0.1328,0.998,B2,uni
0.0274,0.1262,0.0254,0.1277,0.982,D,all
0.0206,0.1285,0.0269,0.136,0.9737,D,mix
0.0174,0.1179,0.0267,0.1316,0.992,D,uni
\end{filecontents*}
\csvautotabular{ds_tab_sc2sc4sc5.csv}
    \caption{Results from scenario B and  D for our Bayesian FBHM with the different priors on $\tau$ for the simulated data from the Weibull distribution. Scenario B1 and B2 is for a treatment effect of $\exp(-0.5)$ and $\exp(-0.275)$ respectively}
\label{Tab:WeiBD_fbhm}
\end{table}

\begin{table}[H]
    \centering
    \begin{filecontents*}{other_tab_sc2a.csv}
bias,beta sd,mse,qv,power,splits,approach
0.2222,0.1171,0.0189,0.081,0.662,3,informed
0.22,0.1149,0.0186,0.0861,0.682,4,informed
0.2276,0.1131,0.0201,0.0917,0.672,5,informed
0.2207,0.1203,0.0188,0.1088,0.652,3,infmix
0.2168,0.1191,0.0186,0.1098,0.658,4,infmix
0.2227,0.1181,0.0195,0.1135,0.654,5,infmix
0.2281,0.1395,0.0245,0.2379,0.476,3,flat
0.2235,0.1394,0.025,0.219,0.484,4,flat
0.2226,0.1396,0.0266,0.2108,0.5,5,flat
0.2312,0.1192,0.0217,0.0777,0.618,3,hierarcal
0.2281,0.1191,0.0204,0.088,0.628,4,hierarcal
0.2275,0.1193,0.0205,0.0985,0.628,5,hierarcal
\end{filecontents*}
\csvautotabular{other_tab_sc2a.csv}
    \caption{Results from scenario B1 with a treatment effect of $\exp{-0.275}$ for the comparison approaches for the simulated data from the Weibull distribution.}
\label{Tab:WeiB1}
\end{table}

\begin{table}[H]
    \centering
    \begin{filecontents*}{other_tab_sc2b.csv}
bias,beta sd,mse,qv,power,splits,approach
-0.0039,0.1214,0.0203,0.0835,0.992,3,informed
-0.0036,0.1194,0.0218,0.0904,0.992,4,informed
0.0064,0.1176,0.0253,0.1019,0.988,5,informed
-0.0045,0.1244,0.0202,0.1103,0.99,3,infmix
-0.0068,0.1233,0.0214,0.1134,0.99,4,infmix
-0.0024,0.1226,0.0239,0.1266,0.99,5,infmix
0.0038,0.1426,0.0266,0.2403,0.95,3,flat
-0.0028,0.1427,0.0279,0.2238,0.954,4,flat
-0.0042,0.1427,0.031,0.2238,0.956,5,flat
0.0061,0.1225,0.0234,0.0802,0.988,3,hierarcal
0,0.1225,0.0226,0.093,0.99,4,hierarcal
-7e-04,0.1226,0.0234,0.1135,0.992,5,hierarcal
\end{filecontents*}
\csvautotabular{other_tab_sc2b.csv}
    \caption{Results from scenario B2 with a treatment effect of $\exp(-0.5)$ for the comparison approaches for the simulated data from the Weibull distribution.}
\label{Tab:WeiB2}
\end{table}

\begin{table}[H]
    \centering
    \begin{filecontents*}{other_tab_sc4.csv}
bias,beta sd,mse,qv,power,splits,approach
0.0727,0.1186,0.0325,0.0811,0.956,3,informed
0.0774,0.1167,0.0305,0.0863,0.954,4,informed
0.0891,0.1148,0.0353,0.0921,0.94,5,informed
0.0724,0.1186,0.0325,0.0821,0.958,3,infmix
0.0771,0.1169,0.0304,0.0871,0.948,4,infmix
0.0877,0.1152,0.0349,0.0932,0.936,5,infmix
0.0039,0.1425,0.0543,0.2402,0.95,3,flat
-0.0044,0.1401,0.035,0.2194,0.944,4,flat
-0.0065,0.1401,0.0405,0.2113,0.95,5,flat
0.0757,0.1204,0.0301,0.0775,0.958,3,hierarcal
0.0711,0.1205,0.0346,0.0881,0.95,4,hierarcal
0.0703,0.1203,0.0379,0.0988,0.954,5,hierarcal
\end{filecontents*}
\csvautotabular{other_tab_sc4.csv}
    \caption{Results from scenario D for the comparison approaches for the simulated data from the Weibull distribution.}
\label{Tab:WeiD}
\end{table}

\subsection{Weibull Mixture Results}

\begin{table}[H]
	\centering
    \begin{filecontents*}{ds_mix_tab_sc1sc3.csv}
bias,beta sd,mse,qv,type1,coverage,scenario,prior
-0.0226,0.1133,0.0376,0.1175,0.0181,0.9819,A,all
-0.0204,0.1193,0.0401,0.1281,0.014,0.986,A,mix
-0.0341,0.1128,0.0475,0.1291,0.02,0.98,A,uni
0.0326,0.1139,0.0457,0.1187,0.0502,0.9498,C,all
0.0292,0.1191,0.0491,0.1298,0.036,0.964,C,mix
0.0238,0.1128,0.0555,0.1306,0.04,0.96,C,uni
\end{filecontents*}
\csvautotabular{ds_mix_tab_sc1sc3.csv}
    \caption{Results from scenario A and scenario C for our Bayesian FBHM with the different priors on $\tau$ for the simulated data from the Weibull mixture distribution.}
\label{Tab:WeiMixAC_fbhm}
\end{table}

\begin{table}[H]
	\centering
    \begin{filecontents*}{mix_other_tab_sc1.csv}
bias,beta sd,mse,qv,type1,coverage,splits,approach
0.0184,0.1119,0.0418,0.0727,0.042,0.958,3,informed
0.0138,0.1094,0.0439,0.0789,0.046,0.954,4,informed
0.0133,0.1075,0.0449,0.0844,0.056,0.944,5,informed
0.0125,0.1156,0.0422,0.0757,0.028,0.972,3,infmix
0.0052,0.1144,0.0447,0.0836,0.034,0.966,4,infmix
0.0025,0.1128,0.0455,0.0898,0.034,0.966,5,infmix
-0.0023,0.1343,0.0519,0.0942,0.02,0.98,3,flat
-0.0023,0.1343,0.0552,0.1043,0.018,0.982,4,flat
-0.0024,0.1345,0.0572,0.1139,0.02,0.98,5,flat
-6e-04,0.114,0.0442,0.0742,0.032,0.968,3,hierarcal
2e-04,0.1158,0.0447,0.0835,0.032,0.968,4,hierarcal
0.0017,0.1174,0.0439,0.0919,0.028,0.972,5,hierarcal
\end{filecontents*}
\csvautotabular{mix_other_tab_sc1.csv}
    \caption{Results from scenario A for the comparison approaches for the simulated data from the Weibull mixture distribution.}
\label{Tab:WeiMixA}
\end{table}

\begin{table}[H]
	\centering
    \begin{filecontents*}{mix_other_tab_sc3.csv}
bias,beta sd,mse,qv,type1,coverage,splits,approach
0.0766,0.1113,0.0445,0.072,0.108,0.892,3,informed
0.0738,0.1088,0.0523,0.078,0.12,0.88,4,informed
0.0784,0.1068,0.0534,0.0832,0.134,0.866,5,informed
0.0604,0.1162,0.0471,0.0763,0.072,0.928,3,infmix
0.055,0.1144,0.0558,0.0834,0.07,0.93,4,infmix
0.056,0.1132,0.0572,0.0902,0.078,0.922,5,infmix
0.0175,0.134,0.0688,0.0939,0.03,0.97,3,flat
0.017,0.1339,0.0809,0.1039,0.028,0.972,4,flat
0.0177,0.1339,0.0808,0.1133,0.03,0.97,5,flat
0.0591,0.1174,0.0707,0.0742,0.078,0.922,3,hierarcal
0.0614,0.1184,0.0622,0.0833,0.074,0.926,4,hierarcal
0.0609,0.1168,0.0618,0.0917,0.074,0.926,5,hierarcal
\end{filecontents*}
\csvautotabular{mix_other_tab_sc3.csv}
    \caption{Results from scenario C for the comparison approaches for the simulated data from the Weibull mixture distribution.}
\label{Tab:WeiMixC}
\end{table}

\begin{table}[H]
	\centering
    \begin{filecontents*}{ds_mix_tab_sc2sc4sc5.csv}
bias,beta sd,mse,qv,power,scenario,prior
0.199,0.1154,0.0387,0.1179,0.73,B1,all
0.203,0.121,0.04,0.1292,0.704,B1,mix
0.1868,0.1148,0.0465,0.1297,0.762,B1,uni
-0.0233,0.1187,0.0389,0.1177,0.99,B2,all
-0.0214,0.1242,0.0404,0.1286,0.988,B2,mix
-0.0391,0.118,0.0454,0.1292,0.992,B2,uni
0.0138,0.1176,0.0417,0.1185,0.986,D,all
0.0086,0.1232,0.043,0.1291,0.988,D,mix
3e-04,0.1167,0.0472,0.1298,0.99,D,uni
\end{filecontents*}
\csvautotabular{ds_mix_tab_sc2sc4sc5.csv}
    \caption{Results from scenario B and scenario D for our Bayesian FBHM with the different priors on $\tau$ for the simulated data from the Weibull mixture distribution. Scenario B1 and B2 is for a treatment effect of $\exp(-0.5)$ and $\exp(-0.275)$ respectively}
\label{Tab:WeiMixBD_fbhm}
\end{table}

\begin{table}[H]
    \centering
    \begin{filecontents*}{mix_other_tab_sc5.csv}
bias,beta sd,mse,qv,power,splits,approach
0.2411,0.1152,0.0491,0.0749,0.622,3,informed
0.2335,0.1131,0.0519,0.082,0.65,4,informed
0.2349,0.1113,0.0486,0.0885,0.662,5,informed
0.2351,0.1179,0.0496,0.0771,0.62,3,infmix
0.2247,0.1168,0.0531,0.0856,0.664,4,infmix
0.2242,0.1158,0.0497,0.0934,0.678,5,infmix
0.2205,0.1361,0.0611,0.0955,0.518,3,flat
0.2199,0.1361,0.0646,0.1063,0.532,4,flat
0.2201,0.1362,0.0608,0.117,0.526,5,flat
0.2206,0.1161,0.0511,0.0754,0.672,3,hierarcal
0.221,0.1162,0.0515,0.0852,0.672,4,hierarcal
0.2226,0.1178,0.0475,0.0947,0.656,5,hierarcal
\end{filecontents*}
\csvautotabular{mix_other_tab_sc5.csv}
    \caption{Results from scenario B1 for the comparison approaches for the simulated data from the Weibull mixture distribution.}
\label{Tab:WeiMixB1}
\end{table}

\begin{table}[H]
    \centering
    \begin{filecontents*}{mix_other_tab_sc2.csv}
bias,beta sd,mse,qv,power,splits,approach
0.0124,0.1192,0.0523,0.0779,0.988,3,informed
0.0093,0.1173,0.0555,0.0866,0.99,4,informed
0.0147,0.1155,0.0522,0.0962,0.99,5,informed
0.0071,0.1221,0.053,0.0805,0.984,3,infmix
0.0028,0.1205,0.0566,0.0899,0.988,4,infmix
0.0049,0.1197,0.0518,0.1021,0.988,5,infmix
-0.0056,0.1391,0.0644,0.0981,0.948,3,flat
-0.0062,0.1391,0.0689,0.1106,0.94,4,flat
-0.0057,0.1392,0.0635,0.1265,0.946,5,flat
-0.0038,0.1193,0.0539,0.0776,0.992,3,hierarcal
-0.0033,0.1192,0.0535,0.0891,0.99,4,hierarcal
-0.003,0.1195,0.0486,0.103,0.992,5,hierarcal
\end{filecontents*}
\csvautotabular{mix_other_tab_sc2.csv}
    \caption{Results from scenario B2 for the comparison approaches for the simulated data from the Weibull mixture distribution.}
\label{Tab:WeiMixB2}
\end{table}

\begin{table}[H]
    \centering
    \begin{filecontents*}{mix_other_tab_sc4.csv}
bias,beta sd,mse,qv,power,splits,approach
-0.0049,0.1376,0.0747,0.0963,0.952,3,informed
0.0507,0.1158,0.0605,0.0838,0.964,4,informed
0.0582,0.114,0.0589,0.0912,0.968,5,informed
0.04,0.1215,0.0599,0.0799,0.974,3,infmix
0.0358,0.1197,0.0619,0.088,0.974,4,infmix
0.0371,0.119,0.0585,0.097,0.978,5,infmix
-0.0049,0.1375,0.0749,0.0963,0.948,3,flat
-0.0062,0.1377,0.0805,0.1077,0.952,4,flat
-0.0056,0.1377,0.0776,0.1193,0.95,5,flat
0.0345,0.1187,0.0928,0.0761,0.976,3,hierarcal
0.0349,0.1187,0.1,0.0865,0.978,4,hierarcal
0.0366,0.1195,0.0932,0.0969,0.972,5,hierarcal
\end{filecontents*}
\csvautotabular{mix_other_tab_sc4.csv}
    \caption{Results from scenario D for the comparison approaches for the simulated data from the Weibull mixture distribution.}
\label{Tab:WeiMixD}
\end{table}

\printbibliography

@article{Escobar1995,
   author = {Michael D. Escobar and Mike West},
   doi = {10.1080/01621459.1995.10476550},
   issn = {1537274X},
   issue = {430},
   journal = {Journal of the American Statistical Association},
   pages = {577-588},
   title = {Bayesian density estimation and inference using mixtures},
   volume = {90},
   year = {1995},
}

@article{Roychoudhury2020,
   author = {Satrajit Roychoudhury and Beat Neuenschwander},
   doi = {10.1002/sim.8456},
   issn = {10970258},
   issue = {7},
   journal = {Statistics in Medicine},
   month = {3},
   pages = {984-995},
   pmid = {31985077},
   publisher = {John Wiley and Sons Ltd},
   title = {Bayesian leveraging of historical control data for a clinical trial with time-to-event endpoint},
   volume = {39},
   year = {2020},
}

@article{Hobbs2013,
   author = {Brian P. Hobbs and Bradley P. Carlin and Daniel J. Sargent},
   doi = {10.1177/1740774513483934},
   issn = {17407745},
   issue = {3},
   journal = {Clinical Trials},
   month = {6},
   pages = {430-440},
   pmid = {23690095},
   title = {Adaptive adjustment of the randomization ratio using historical control data},
   volume = {10},
   year = {2013},
}

@article{Han2017,
   author = {Baoguang Han and Jia Zhan and Z. John Zhong and Dawei Liu and Stacy Lindborg},
   doi = {10.1002/pst.1815},
   issn = {15391612},
   issue = {4},
   journal = {Pharmaceutical Statistics},
   month = {7},
   pages = {296-308},
   pmid = {28560815},
   publisher = {John Wiley and Sons Ltd},
   title = {Covariate-adjusted borrowing of historical control data in randomized clinical trials},
   volume = {16},
   year = {2017},
}

@article{Besag1995,
   author = {Julian Besag and Charles Kooperberg},
   isbn = {201401:54:06},
   issue = {4},
   journal = {Biometrika},
   pages = {733-779},
   title = {On conditional and intrinsic autoregressions},
   volume = {82},
   year = {1995},
}

@article{Green1995,
   author = {Peter J Green},
   issue = {4},
   journal = {Biometrika},
   pages = {711-743},
   title = {Reversible jump Markov chain Monte Carlo computation and Bayesian model determination},
   volume = {82},
   year = {1995},
}

@article{Viele2014,
   author = {Kert Viele and Scott Berry and Beat Neuenschwander and Billy Amzal and Fang Chen and Nathan Enas and Brian Hobbs and Joseph G. Ibrahim and Nelson Kinnersley and Stacy Lindborg and Sandrine Micallef and Satrajit Roychoudhury and Laura Thompson},
   doi = {10.1002/pst.1589},
   issn = {15391612},
   issue = {1},
   journal = {Pharmaceutical statistics},
   pages = {41-54},
   pmid = {23913901},
   title = {Use of historical control data for assessing treatment effects in clinical trials.},
   volume = {13},
   year = {2014},
}

@article{Schmidli2014,
   author = {Heinz Schmidli and Sandro Gsteiger and Satrajit Roychoudhury and Anthony O'Hagan and David Spiegelhalter and Beat Neuenschwander},
   doi = {10.1111/biom.12242},
   issn = {15410420},
   issue = {4},
   journal = {Biometrics},
   month = {12},
   pages = {1023-1032},
   pmid = {25355546},
   title = {Robust meta-analytic-predictive priors in clinical trials with historical control information},
   volume = {70},
   year = {2014},
}

@article{Su2022,
   author = {Liwen Su and Xin Chen and Jingyi Zhang and Fangrong Yan},
   issue = {e2100394},
   journal = {Precision Oncology},
   title = {Comparative study of Bayesian information borrowing methods in oncology clinical trials},
   volume = {6},
   url = {https://doi.org/10.},
   year = {2022},
}

@article{Savic2017,
   author = {Margaret Gamalo-Siebers and Jasmina Savic and Cynthia Basu and Xin Zhao and Mathangi Gopalakrishnan and Aijun Gao and Guochen Song and Simin Baygani and Laura Thompson and H. Amy Xia and Karen Price and Ram Tiwari and Bradley P. Carlin},
   doi = {10.1002/pst.1807},
   issn = {15391612},
   issue = {4},
   journal = {Pharmaceutical Statistics},
   month = {7},
   pages = {232-249},
   pmid = {28448684},
   publisher = {John Wiley and Sons Ltd},
   title = {Statistical modeling for Bayesian extrapolation of adult clinical trial information in pediatric drug evaluation},
   volume = {16},
   year = {2017},
}

@article{Murray2016,
   author = {Thomas A. Murray and Brian P. Hobbs and Daniel J. Sargent and Bradley P. Carlin},
   doi = {10.1214/15-BA954},
   issn = {19316690},
   issue = {2},
   journal = {Bayesian Analysis},
   month = {6},
   pages = {381-402},
   publisher = {International Society for Bayesian Analysis},
   title = {Flexible Bayesian survival modeling with semiparametric time-dependent and shape-restricted covariate effects},
   volume = {11},
   year = {2016},
}

@article{Hobbs2011,
   author = {Brian P. Hobbs and Bradley P. Carlin and Sumithra J. Mandrekar and Daniel J. Sargent},
   doi = {10.1111/j.1541-0420.2011.01564.x},
   issn = {15410420},
   issue = {3},
   journal = {Biometrics},
   pages = {1047-1056},
   title = {Hierarchical commensurate and power prior models for adaptive incorporation of historical information in clinical trials},
   volume = {67},
   year = {2011},
}

@article{Neuenschwander2010,
   author = {Beat Neuenschwander and Gorana Capkun-Niggli and Michael Branson and David J. Spiegelhalter},
   doi = {10.1177/1740774509356002},
   issn = {17407745},
   issue = {1},
   journal = {Clinical Trials},
   month = {2},
   pages = {5-18},
   pmid = {20156954},
   title = {Summarizing historical information on controls in clinical trials},
   volume = {7},
   year = {2010},
}

@article{Bi2023,
   author = {Dehua Bi and Meizi Liu and Jianchang Lin and Rachael Liu},
   doi = {10.1002/sim.9936},
   issn = {10970258},
   journal = {Statistics in Medicine},
   publisher = {John Wiley and Sons Ltd},
   title = {BEATS: Bayesian hybrid design with flexible sample size adaptation for time-to-event endpoints},
   year = {2023},
}

@article{Hupf2021,
   author = {Bradley Hupf and Veronica Bunn and Jianchang Lin and Cheng Dong},
   doi = {10.1002/sim.8970},
   issn = {10970258},
   issue = {14},
   journal = {Statistics in Medicine},
   month = {6},
   pages = {3385-3399},
   pmid = {33851441},
   publisher = {John Wiley and Sons Ltd},
   title = {Bayesian semiparametric meta-analytic-predictive prior for historical control borrowing in clinical trials},
   volume = {40},
   year = {2021},
}

@article{Daniel2021,
   author = {Rhian Daniel and Jingjing Zhang and Daniel Farewell},
   doi = {10.1002/bimj.201900297},
   issn = {15214036},
   issue = {3},
   journal = {Biometrical Journal},
   month = {3},
   pages = {528-557},
   pmid = {33314251},
   publisher = {John Wiley and Sons Inc},
   title = {Making apples from oranges: Comparing noncollapsible effect estimators and their standard errors after adjustment for different covariate sets},
   volume = {63},
   year = {2021},
}

@article{Ibrahim2015,
   author = {Joseph G. Ibrahim and Ming Hui Chen and Yeongjin Gwon and Fang Chen},
   doi = {10.1002/sim.6728},
   issn = {10970258},
   issue = {28},
   journal = {Statistics in Medicine},
   month = {12},
   pages = {3724-3749},
   pmid = {26346180},
   publisher = {John Wiley and Sons Ltd},
   title = {The power prior: Theory and applications},
   volume = {34},
   year = {2015},
}

@article{Lewis2019,
   author = {Connor Jo Lewis and Somnath Sarkar and Jiawen Zhu and Bradley P. Carlin},
   doi = {10.1080/19466315.2018.1497533},
   issn = {19466315},
   issue = {1},
   journal = {Statistics in Biopharmaceutical Research},
   month = {1},
   pages = {67-78},
   publisher = {Taylor and Francis Inc.},
   title = {Borrowing From Historical Control Data in Cancer Drug Development: A Cautionary Tale and Practical Guidelines},
   volume = {11},
   year = {2019},
}

@article{FitzGerald2016,
   author = {J. Mark FitzGerald and Eugene R. Bleecker and Parameswaran Nair and Stephanie Korn and Ken Ohta and Marek Lommatzsch and Gary T. Ferguson and William W. Busse and Peter Barker and Stephanie Sproule and Geoffrey Gilmartin and Viktoria Werkström and Magnus Aurivillius and Mitchell Goldman},
   doi = {10.1016/S0140-6736(16)31322-8},
   issn = {1474547X},
   issue = {10056},
   journal = {The Lancet},
   month = {10},
   pages = {2128-2141},
   pmid = {27609406},
   publisher = {Lancet Publishing Group},
   title = {Benralizumab, an anti-interleukin-5 receptor alpha monoclonal antibody, as add-on treatment for patients with severe, uncontrolled, eosinophilic asthma (CALIMA): a randomised, double-blind, placebo-controlled phase 3 trial},
   volume = {388},
   year = {2016},
}

@article{Bleecker2016,
   author = {Eugene R. Bleecker and J. Mark FitzGerald and Pascal Chanez and Alberto Papi and Steven F. Weinstein and Peter Barker and Stephanie Sproule and Geoffrey Gilmartin and Magnus Aurivillius and Viktoria Werkström and Mitchell Goldman},
   doi = {10.1016/S0140-6736(16)31324-1},
   issn = {1474547X},
   issue = {10056},
   journal = {The Lancet},
   month = {10},
   pages = {2115-2127},
   pmid = {27609408},
   publisher = {Lancet Publishing Group},
   title = {Efficacy and safety of benralizumab for patients with severe asthma uncontrolled with high-dosage inhaled corticosteroids and long-acting beta2-agonists (SIROCCO): a randomised, multicentre, placebo-controlled phase 3 trial},
   volume = {388},
   year = {2016},
}

@article{Neuenschwander2016,
   author = {Beat Neuenschwander and Simon Wandel and Satrajit Roychoudhury and Stuart Bailey},
   doi = {10.1002/pst.1730},
   issn = {15391612},
   issue = {2},
   journal = {Pharmaceutical Statistics},
   month = {3},
   pages = {123-134},
   pmid = {26685103},
   publisher = {John Wiley and Sons Ltd},
   title = {Robust exchangeability designs for early phase clinical trials with multiple strata},
   volume = {15},
   year = {2016},
}

\pagebreak

\setcounter{equation}{0}
\setcounter{figure}{0}
\setcounter{table}{0}
\setcounter{section}{0}
\setcounter{page}{0}

\title{Supplementary Material - Borrowing from historical control data in a Bayesian time-to-event model with flexible baseline hazard function}
\maketitle

\begin{refsection}

\section{Truncated Poisson}
The prior for the number of split points in the piecewise exponential likelihood is a right-truncated Poisson. Defining the normalising constant as
\begin{equation}
    q_t =  \sum_{j=0}^{J_{\text{max}}} \frac{e^{-\phi}\phi^j}{j!} ,
\end{equation}
the probability density function is 
\begin{equation}
    p_J(j) = 
    \begin{cases} 
            \frac{e^{-\phi}\phi^j}
            {j!} \frac{1}{q_t} & \text{if } j \text{ is } \leq J_{\text{max}}\\ 
        0 & \text{if } j \text{ is } > J_{\text{max}} 
    \end{cases}.
\end{equation}
The  expectation is 
\begin{equation}
     \mathbb{E}(J) = \phi \frac{q_{t-1}}{q_t},
\end{equation}
where the sum  of the pdf over 0 to $J_{\text{max}} -1$ is defined as 
\begin{equation}
    q_{t-1} = \sum_{j=0}^{J_{\text{max}}-1} \frac{e^{-\phi}\phi^j}{j!}.
\end{equation}

The variance is 
\begin{equation}
    \mathbb{V}(J) = \phi^2 \left( \frac{q_{t-2}q_t-q_{t-1}}{qt} \right)
\end{equation}

\section{Gaussian conditional autoregression}

Gaussian conditional autoregressions with a Markov property are also known as Gaussian Markov random fields. If we begin with the multivariate normal 
\begin{equation*}
    \pi(\vec{x}) = \frac{|Q|^{1/2}}{(2\pi)^{\frac{n}{2}}}\exp\left(
    -\frac{1}{2} (\vec{x}- \bm{\mu})'Q(\vec{x}- \bm{\mu})
    \right)
\end{equation*}
where $Q$ is the precision. The conditional distribution is normal with mean and variance
\begin{align*}
    \mathbb{E}(x_i|\vec{x}_{-i}) & = \mu_i + \sum_{j\neq i}\gamma_{ij}(x_j - \mu_j) \\
    \text{Var}(x_i|\vec{x}_{-i}) &= \kappa_i^{-1}
\end{align*}
Under
the assumption that
$\kappa_i \gamma_{ij} = \kappa_j \gamma_{ji}$ for all $i = j$, these conditional distributions correspond to a multivariate joint Gaussian
distribution with mean 0 and precision matrix $Q$ with elements $Q_{ii} = \kappa_{i}$ and $Q_{ij} = -\kappa_i\gamma_{ij}$ ,
$i \neq j$, provided that $Q$ is symmetric and positive definite.

A common approach is to specify the joint density of a zero-mean Gaussian Markov random field implicitly by specifying each of the $n$ full conditionals. 

In our prior structure for the log baseline hazards, we are motivated by two key aspects. The baseline hazards are unlikely to be independent; rather, we link the baseline hazard with the adjacent hazards, smoothing their value. The variance is inversely related to the length of the interval; the larger the interval, the greater the information, the smaller the variance. Motivated by this, we define the length of the $I_j$ interval as $\hat{\Delta} = s_j - s_{j-1}$. We make $\beta_{ij}$ non-zero for $i \sim j$ (adjacent time intervals $I_{j+1}$ and $I_{j-1}$), its size proportional to the length of the time interval. We scale the conditional variance inversely by the sum of the corresponding interval (multiplied by 2) and adjacent interval. 

\section{Gaussian Markov Random Field Prior}
 
For the prior on the historical log hazard one could assign independent priors to each of the $J+1$ components of $\bm{\lambda}_0$. However, as $\bm{\lambda}_0(\cdot)$ is likely a smooth function over time the components of $\bm{\lambda}_0$ are unlikely to be independent of each other a priori. We specify the prior for the components of the historical $\bm{\lambda}_0$ as a Gaussian Markov random field with a nearest-neighbour structure in the form $\log(\lambda_{0j})|\left\{\log(\lambda_{0k}), j \neq k \right\} \sim N(\nu_j, \sigma^2_j)$

\begin{equation}
\label{eq:ICAR_nu}
    \nu_j = \mu + \sum _{k \neq j} W_{jk}(\log(\lambda_{0k})- \mu)
\end{equation}
where the hyperparameter $\mu = \mathbb{E}(\log(\lambda_j))$ represents the overall trend in the the levels of the log hazard function and $W_{j(j-1)} = l_j$ and $W_{j(j+1)}= r_j$ are the influences of the left and right neighbours of $\log(\lambda_j)$ respectively. All other $W_{jk}$ where $k \notin \{j-1, j, j+1 \}$ are set to 0. 

To guarantee that the joint distribution is Gaussian, the following conditions with respect to the conditional distribution need to be satisfied, $s_j$ and $r_j$ are non-negative with $l_j + r_j\leq 1$ and $l_j \sigma^2_{j+1} = r_{j+1} \sigma^2_j$. We specify $l_j$, $r_j$ and $\sigma^2_j$, with the aim of forcing the corresponding hazard function to be ``smooth". Of the two neighbours of $\lambda_{0j}$, the one corresponding to the longer interval should have the greatest influence on the mean $\nu_j$. 

Letting $\bar{\Delta}_j= s_j-s_{j-1}$ denote the length of the $I_j$ interval, the weights for the intervals adjacent to the $j$ interval are
\begin{align}
    l_j =& \frac{c_\lambda (\bar{\Delta}_{j-1} + \bar{\Delta}_j)}
    {\bar{\Delta}_{j-1} + 2\bar{\Delta}_j + \bar{\Delta}_{j+1}},\\
     r_j =& \frac{c_\lambda(\bar{\Delta}_{j+1} + \bar{\Delta}_j)}
    {\bar{\Delta}_{j-1} + 2\bar{\Delta}_j + \bar{\Delta}_{j+1}},
\end{align}
and the remaining non-adjacent weights are set to zero. The level of dependence between adjacent intervals is controlled by $c_\lambda~\in~[0,1]$. The conditional variance is also a function of the split times,   $\sigma^2_j=\sigma^2_\lambda Q_j$ where $\sigma^2_\lambda$ is an overall measure of variation across the $\lambda_j$'s and 
\begin{equation}
\label{eq:ICAR_Q}
    Q_j = \frac{2}{\bar{\Delta}_{j-1} + 2\bar{\Delta}_j + \bar{\Delta}_{j+1}}.
\end{equation}

This gives us an overall precision matrix $\Omega$ with elements $\Omega_{ii} = 1 / \sigma^2_i$ and $\Omega_{ij} =  -(1 / \sigma^2_i) W_{ij}$. We can thus obtain the precision matrix by $\Omega = M^{-1}(I - C)$ where $C=(c_{ij})_{1\leq i,j\leq m}$, $c_{ii+1} = r_i$ and $c_{ii-1} = l_i$  and $M = \text{diag}(\sigma^2_1, ..., \sigma^2_{J+1})$.

\section{Summary of the algorithm}

We first describe the likelihood in detail and then provide a brief summary of the updates which are performed in the algorithm. 

\subsection{Likelihood}
The likelihood for the current trial is the following, consider partitioning into $J+1$ intervals with split points $0=s_0 <s_1 <...<s_{J+1}$ with $s_{J+1} > y_i$ for all $i=1,2,...,n$. Thus, we have $J$ partitions and $J+1$ intervals $(0, s_1],...(s_{J-1}, s_{J+1}]$. In the $j$th interval, we assume a constant baseline hazard $h_{j}(y_i) = \lambda_{j}$ for $y_i \in I_j = (s_j-1,s_j]$· In order to ensure that the algorithm is stable $s_{J+1}$ must be set to $\text{max}(y_i|\nu_i =1)$, the largest observed non-censoring time.

Let $\vec{D} =(n,\vec{y},\vec{X},\bm{\nu})$ denote the observed data with $\nu_i=1$ if the $i$th subject failed (death or event happens) and 0 otherwise, and $\vec{X}$ is the $n \times p$ matrix
of covariates with $i$th row $\vec{x}'_i$. The likelihood function can be expressed in terms of the baseline hazard and survival function $(S(y_i ;\bm{\lambda},\bm{\beta}, \vec{x}_i))$ as 
\begin{align}
    \label{eq:like_basic}
    \mathcal{L}(\bm{\beta}, \bm{\lambda}, \vec{s} |\vec{D}) & =
    \prod_{i=1}^n\prod_{j=1}^{J+1} 
    (h_j(y_i)  \exp\left\{\vec{x}_i'\bm{\beta}\right\})^{\delta_{ij}\nu_i}S(y_i ;\bm{\lambda},\bm{\beta}, \vec{x}_i), 
\end{align}
where $\delta_{ij} = 1$ if the $i$th subject failed or was censored in the $j$th interval, and
0 otherwise, $\bm{\lambda} = (\lambda_1, ..., \lambda_J)$ is the vector of baseline hazards, $\vec{x}_i' = (x_{i1} , x_{i2}, ... , x_{ip})$ denotes the $p \times 1$ vector of covariates
for the $i$th subject, and $\bm{\beta} = (\beta_1, ..., \beta_p)$ is the corresponding vector of regression coefficients. Expressing the survival function in terms of the cumulative hazard gives 
\begin{align}
    \mathcal{L}(\bm{\beta}, \bm{\lambda}, \vec{s} |\vec{D}) =
           & \prod_{i=1}^n\prod_{j=1}^{J+1} (\lambda_j \exp\left\{\vec{x}_i'\bm{\beta}\right\})^{\delta_{ij}\nu_i}
     \nonumber \exp \Big\{
    -\delta_{ij} \Big(
    \lambda_j(y_i-s_{j-1})\\
    & \qquad  + \sum_{g=1}^{j-1}\lambda_g (s_g-s_{g-1})
    \Big)\exp\{\vec{x}_i'\bm{\beta}
    \}
    \Big\}
\end{align}

We can simplify this by setting $\nu_{ij}= \delta_{ij} \nu_i$ which is 1 if the subject failed in the $j$th interval and 0 otherwise and letting $t_{ij}$ denote the observed event time in $[s_{j-1}, s_j)$,
\begin{equation}
    t_{ij} = 
    \begin{cases}
    s_j-s_{j-1} & \text{if $y_{i} \geq s_j$},\\
    y_{i} - s_{j-1} & \text{if $y_{i} \in [s_{j-1}, s_j)$},\\
    0 & \text{otherwise}.
  \end{cases} 
\end{equation}

Thus we can express the log likelihood per $j$th time point as 
\begin{align}
    \log  \mathcal{L}_j(\bm{\beta}, \bm{\lambda}, \vec{s} |\vec{y}, \vec{X}) = &  d_j \log(\lambda_j) + \sum_i \nu_{ij} \vec{x}_i'\bm{\beta} -
    \lambda_j \sum_i
     t_{ij}\exp\{\vec{x}_i'\bm{\beta}
    \},
\end{align}
where $\sum_i \nu_{ij} = d_j$ is the number of events in interval $\vec{I}_j = (s_{j-1}, s_j]$ and $\sum_i t_{ij}$ is the total time to an event or last follow up in the $j$th time interval. 

\subsection{Updates}

In this section we describe a summary of the posterior updates which allow us to obtain samples from the joint posterior.  The joint posterior in full 
\begin{align*}
\pi(J)\pi(\bm{s}|J)\mathcal{L}(\cdot|\vec{D})\pi(\bm{\beta})\pi(\bm{\tau}|\vec{s})\pi(\bm{\lambda}|\bm{\lambda}_0,\bm{\tau}, \vec{s}) \mathcal{L}(\cdot|\vec{D}_0)\pi(\bm{\beta}_0) \pi(\bm{\lambda}_0|\vec{s})\pi(\mu)\pi(\sigma^{2}),
\end{align*}
leads to the following conditionally conjugate updates which can be sampled via a Gibbs move  
\begin{align*}
    \mu|\lambda_0,  \sigma^2, J, \vec{s}  \sim&  \mathcal{N}_{J+1}\left(
    \frac{\vec{1}'(\Sigma_s)^{-1} \log(\bm{\lambda}_0)}
    {\vec{1}'(\Sigma_s)^{-1} \vec{1}},
    \frac{\sigma^2}{\vec{1}'(\Sigma_s)^{-1} \vec{1}}
    \right), \\
    \sigma^2_\lambda | \mu, \lambda_0, J, \vec{s}
     \sim &  \text{IG} \left( 
    \frac{(J+1)}{2} + a_\sigma, \frac{(\mu \vec{1} - \log(\bm{\lambda}_0))'\Sigma_s^{-1}  
    (\mu \vec{1} - \log(\bm{\lambda}_0))}{2} + b_\sigma
    \right) .
\end{align*}

The updates for each of the commensurate priors are conjugate. The mixture prior is of the form  
\begin{align}
    \tau_j^{(\text{mix})}|\cdot  \sim&  (q)  \text{IG}\left(\frac{1}{2} +a_\tau, \frac{(\log(\lambda_j)-\log(\lambda_{0j}))^2}{2} +b_\tau\right) +  \nonumber\\
    & (1-q) \text{IG} \left(\frac{1}{2}+c_\tau, \frac{(\log(\lambda_j)-\log(\lambda_{0j}))^2}{2} +d_\tau\right)
\end{align}
where the weights are proportional to the product of the original weights and the marginal likelihood with respect to the prior. Given the number of normalising constants 
\begin{align}
w_0 &= \frac{ b_\tau^{a_\tau}\Gamma(a_\tau + \frac{1}{2})}{\sqrt{2\pi}(\frac{1}{2}(\log(\lambda_j) - \log(\lambda_{0j}))^2)+b_\tau)^{\frac{1}{2}+a_\tau}\Gamma(a_\tau) },\\
w_1 & =  \frac{ d_\tau^{c_\tau}\Gamma(c_\tau + \frac{1}{2})}{\sqrt{2\pi}(\frac{1}{2}(\log(\lambda_j) - \log(\lambda_{0j}))^2)+d_\tau)^{\frac{1}{2}+c_\tau}\Gamma(c_\tau) }
\end{align}
$w = p_0w_0 + p_1w_1$, 
$q = \frac{p_0w_0}{w}$ and $(1- q) = \frac{p_1w_1}{w}$. 

Here we summarise how we sample the remaining parameters $J, \vec{s}, \bm{\beta}_0, \bm{\beta}, \bm{\lambda}_0$ and $\bm{\lambda}$  via  a series of Metropolis-Hasting (MH) steps, the full derivations are below. 

To sample the total number of splits $J$ a change in the dimension of the parameter space is required. A reversible jump MH move is included which either adds or deletes a split point, (or proposes a global change to $J$ with a whole new set of split points). For a birth move a random split point is sampled from $s^* \sim U(0,s_{J+1})$ say between $s_{j-1}$ and $s_{j}$, and proposals for the new baseline hazards and corresponding between hazard variability for $(s^*, s_j]$ and $(s_{j-1}, s^*]$ are made. In order to ensure the transformation from the current parameter values and random variables for the proposed move to the new state to be a diffeomorphism, we require the dimensions to match. The corresponding baseline hazards (historical and current) and variances $(\tau_j^*, \tau^*_{j+1})$ are proposed from two random perturbations of the form $\frac{x_{j+1}^*}{x_{j}^*} = \frac{1-U}{U}$ where $U$ is a standard uniform random variable. This allows us to propose two model parameters under the constraint of a single proposal random variable. The death move starts with proposing to delete a time split $s_1,...,s_J$  and the adjusted baseline hazard and variances  are proposed via a weighted mean on the log scale. 

We sample the split points $\vec{s}$ by shuffling $s_1,...,s_J$ (as $s_0$ and $s_{J+1}$ are fixed). For each split $s_j$ ($j=1,...,J)$ propose a new time split from a continuous uniform distribution $s^* \sim U(s_{j-1}, s_{j+1})$. 

To update each element of the baseline hazard $\lambda_{0j}$ and $\lambda_j$ for all $j=1,...,J$ we exploit the conjugacy of the likelihood with a Gamma prior, to propose from a conditional conjugate posterior under the assumption of independence.   

In the case of the historic data we have 
\begin{equation}
    \lambda_{0j} | \bm{\beta}_0, \vec{D}_0 \sim G\left(a_\lambda + d_{0j}, b_\lambda + \sum_i t_{0ij} \exp \left\{
    \vec{x}_{0i}'\bm{\beta}_0
    \right\} \right)
\end{equation}
where $\delta_{ij}$ identifies the  set  of patients with an event or censored in the interval $(s_{j-1}, s_j]$ and $d_{0j}$ is the number of historic events in interval $(s_{j-1}, s_j]$ and $t_{0ij}$ is the time to an event or last follow up in the $j$th time interval. We set $a_\lambda = 0.01$ and $b_\lambda = 0.01$ to be vague. This simplifies the algorithm as we avoid a proposal tuning parameter. For the current baseline hazard we add a power parameter on the historical likelihood 
\begin{equation*}
\mathcal{L}(\bm{\beta}_0, \bm{\lambda}_0, \vec{s}|\vec{D})^{\alpha}
\end{equation*}
with $\alpha = 0.35$ to down weight the information. This gives us a proposal for the current data of   
\begin{equation}
    \lambda_j |\bm{\beta}, \vec{D} \sim G\left(a_\lambda +  d_j +  \alpha d_{0j}, b_\lambda + \sum_{i=1}^n  t_{0ij} \exp \left\{
    \vec{x}_i'\bm{\beta}
    \right\}
    + \alpha \sum_{i=1}^{n_0}  t_{ij} \exp \left\{
    \vec{x}_{0i}'\bm{\beta}_0
    \right\}
    \right)
\end{equation}

We sample each element of the regression coefficients $\beta_{0k}$ and $\beta_l$ for all $k=1,...,p_0$ and $l=1,...,p$ via a Metropolis adjusted Langevin algorithm with proposal variance $c_\beta$ which requires tuning.

\section{Metropolis Hastings updates}
Here we describe the updates which require a Metropolis Hastings step in full. As the first and second derivatives of the target density are analytically tractable, we use a Newton-Raphson iteration. This allows to have just one scaling factor for each vector of regression
coefficients $\bm{\beta}$ and $\bm{\beta}_0$.

\subsection{MH beta moves}
The full conditional distribution for $\bm{\beta}_0$ is 
\begin{align*}
    \pi(\bm{\beta}_0|\vec{D}_0,\vec{s}, \bm{\lambda}_0) \propto& L(\vec{D}_0)\pi(\bm{\beta}_0) \\
     \propto &\prod_{i=1}^{n_0}\prod_{j=1}^{J+1} (\lambda_{0j} \exp\left\{\vec{x}_{i0}'\bm{\beta}_0\right\})^{\delta_{ij}\nu_i}
     \nonumber \exp \Big\{
    -\delta_{ij} \Big(
    \lambda_{0j}(y_{i0}-s_{j-1}) + \\
    & \qquad  + \sum_{g=1}^{j-1}\lambda_{0g} (s_g-s_{g-1})
    \Big)\exp\{\vec{x}_{i0}'\bm{\beta}_0
    \}
    \Big\}
\end{align*}

The coefficients $\bm{\beta}_{0k}$ for all $k \in  \{1, ..., p\}$ are updated as follows. The first $D_1(\beta_{0k})$ and
second $D_2(\beta_{0k})$ derivatives of the log conditional posterior are given by
\begin{align*}
    D_1(\beta_{0k}) =& \sum_i\sum_j (\delta_{ij}\nu_i) x_{0ik} - \delta_{ij}\left(\lambda_{0j}(y_{0i}-s_{j-1}) +  \sum_{g=1}^{j-1}\lambda_{0g}(s_g -s_{g-1})\right) x_{0ik}\exp\left\{
    \vec{x}_{0i}'\bm{\beta}_{0}
    \right\}, \\
    D_2(\beta_{0k}) =& -\sum_i\sum_j   \delta_{ij}\left(\lambda_{0j}(y_{0i}-s_{j-1}) +  \sum_{g=1}^{j-1}\lambda_{0g}(s_g -s_{g-1})\right) x^2_{0ik}\exp\left\{
    \vec{x}_{0i}'\bm{\beta}_{0}
    \right\} .\\
\end{align*}
It is noted that the conditional posterior of $\beta_{0k}$ is a log-concave function because $D(\beta_{0k}) < 0$. We approximate the target density by a normal distribution with mean $m_{\beta_0}(\beta_{0k}) =\beta_{0k} -D_1(\beta_{0k})/D_2(\beta_{0k})$ 
and variance 
$V_{\beta_{0k}}(\beta_{0k}) = - c^2_{\beta_{0k}} / D_2(\beta_{0k})$ for our proposal. We set
$c_{\beta_0}$ to achieve an acceptance rate of $40\% \sim 45\%$.

Propose
\begin{equation}
    \label{eq:AL_beta}
    \beta_{0k}^* \sim N( m_{\beta_0}(\beta_{0k}^{(\text{old})}), V_{\beta_{0k}}(\beta_{0k}^{(\text{old})}))
\end{equation}
which gives an acceptance probability of 
\begin{align*}
    \text{Target Ratio} = \frac{\pi (\beta_{0k}^{*}|\bm{\beta}_{0k}^{-k}, \vec{s}, \bm{\lambda}_0, \vec{D}_0)  }{
    \pi (\beta_{0k}^{(\text{old})}|\bm{\beta}_{0k}^{-k}, \vec{s}, \bm{\lambda}_0, \vec{D}_0) 
    }\\
    \text{Prop Ratio} = 
   \frac{N(\beta_{0k}^{(\text{old})}| \text{m}_{\beta_{0k}}(\beta_{0k}^{*}),V_{\beta_{0k}}(\beta_{0k}))}{
   N(\beta_{0k}^{*}| \text{m}_{\beta_{0k}}(\beta_{0k}^{(\text{old})}), V_{\beta_{0k}}(\beta_{0k}^{(\text{old})}))
   }.
\end{align*}

\subsection{MH baseline hazard}
The full conditional distribution for $\lambda_{0j}$ is 
\begin{align}
    L(\cdot|\vec{D}_0)\pi(\log(\bm{\lambda}_0)) \propto & \prod_{i=1}^n\prod_{j=1}^{J+1} (\lambda_{0j} \exp\left\{\vec{x}_{0i}'\bm{\beta}_0\right\})^{\delta_{ij}\nu_i}
     \nonumber \exp \Big\{
    -\delta_{ij} \Big(
    \lambda_{0j}(y_{0i}-s_{j-1})\\
    & \qquad  + \sum_{g=1}^{j-1}\lambda_{0g} (s_g-s_{g-1})
    \Big)\exp\{\vec{x}_i'\bm{\beta}_0
    \}
    \Big\} \pi(\log(\bm{\lambda}_0)) 
\end{align}

We exploit conjugacy with the likelihood from a independent Gamma prior with initial parameters $a_\lambda, b_\lambda$ for our proposal. The conditional posterior for $\lambda_{0j}$ is dependent on the value of $j$. When $j=1$, $\lambda_1$ appears in each iteration of $j$, either in the exponent for $j=1$
\begin{equation}
    \exp \left\{ -\sum _i \delta_{i1}(y_{0i} - s_{0})\exp \left\{
    \vec{x}_{0i}'\bm{\beta}_0
    \right\} \right\}
\end{equation}
or in the form of $\lambda_{01}(s_1 - s_0)$ for $j=2,...,J+1$. Let $t_{ij}$ denote the observed event time in $[s_{j-1}, s_j)$,
\begin{equation}
    t_{ij} = 
    \begin{cases}
    s_j-s_{j-1} & \text{if $y_{0i} \geq s_j$},\\
    y_{0i} - s_{j-1} & \text{if $y_{0i} \in [s_{j-1}, s_j)$},\\
    0 & \text{otherwise}.
  \end{cases} 
\end{equation}
The conjugate update for the baseline hazard from the historic data is 
\begin{equation}
    \lambda_{0j} | \vec{D}_0 \sim G\left(a_\lambda + \sum_i d_{0j}, b_\lambda + \sum_i t_{ij}\exp \left\{
    \vec{x}_{0i}'\bm{\beta}_0
    \right\} \right)
\end{equation}
where $d_j$ is the number of events in interval $(s_{j-1}, s_j]$ and $\sum_i t_{ij}$ is the total time to an event or last follow up in the $j$th time interval. We set $a_\lambda = 0.01$ and $b_\lambda = 0.01$. 

We are proposing on the $\lambda_0$ scale, but our prior is on the $\log(\lambda_0)$ scale, a Jacobian is included in the prior ratio from the change of variable. Our probability of acceptance in the target ratio includes the Jacobian from the change of variable. This gives  
\begin{align*}
    \text{Target ratio} =&  \frac{\mathcal{L}(\bm{\beta}_0,\bm{\lambda}^{*}_0, \vec{s}, J|\vec{D}_0)\pi(\lambda_{j}^{*}|\lambda_{0j}^{*},\tau_j, \vec{s})\pi(\lambda_{0j}^{*}|\bm{\lambda}_{0}^{(-j)},\vec{s})
    \frac{1}{\lambda_{0j}^*}
    }{
    \mathcal{L}(\bm{\beta}_0,\bm{\lambda}^{(\text{old})}_0, \vec{s}, J|\vec{D}_0)\pi(\lambda_{j}|\lambda_{0j}^{(\text{old})},\tau_j,\vec{s})\pi(\lambda_{0j}^{(\text{old}
    )}|\bm{\lambda}_{0}^{(-j)},\vec{s}) \frac{1}{\lambda_{0j}^{(\text{old})}}
    } \\
    \text{Proposal ratio} = &  \frac{G(\lambda_{0j}^{(\text{old})}| a^*_{0j\lambda}, b^*_{0j\lambda})
    }{
    G(\lambda_{0j}^{*}| a^*_{0j\lambda}, b^*_{0j\lambda})
    }
\end{align*}

For the current baseline hazard a power parameter on the historical likelihood 
\begin{equation*}
\mathcal{L}(\bm{\beta}_0, \bm{\lambda}_0, \vec{s}|\vec{D})^{\alpha}
\end{equation*}
with $\alpha = 0.3$ to down weight it. This gives us a proposal $ \lambda_j | \bm{\lambda}^{(-j)}$ of   
\begin{equation*}
    G\left(a_\lambda +  d_j +  \alpha d_{0j}, b_\lambda + \sum_{i=1}^n \delta_{ij}(y_i - s_{j-1})\exp \left\{
    \vec{x}_i'\bm{\beta}
    \right\}
    + \alpha \sum_{i=1}^{n_0}  \delta_{ij}(y_{0i} - s_{j-1})\exp \left\{
    \vec{x}_{0i}'\bm{\beta}_0
    \right\}
    \right)
\end{equation*}

with a probability of acceptance of 

\begin{align}
    \text{Target ratio} =&  \frac{\mathcal{L}(\bm{\beta},\bm{\lambda}^{*}, \vec{s}, J|\vec{D})\pi(\lambda_{j}^{*}|\lambda_{0j}^{*},\tau_j, \vec{s})
    \frac{1}{\lambda_{j}^*}
    }{
    \mathcal{L}(\bm{\beta},\bm{\lambda}^{(\text{old})}, \vec{s}, J|\vec{D})\pi(\lambda^{(\text{old})}_{j}|\lambda_{0j},\tau_j,\vec{s}) \frac{1}{\lambda_{j}}
    } \label{eq:proplam}\\
    \text{Proposal ratio} = &  \frac{G(\lambda_{j}^{(\text{old})}| a^*_{j\lambda}, b^*_{j\lambda})
    }{
    G(\lambda_{j}^{*}| a^*_{j\lambda}, b^*_{j\lambda})
    } . \nonumber 
\end{align}

\subsection{Shift locations move}
Sample the locations of $\vec{s}$ via a Metropolis-Hastings move that shifts the time splits $s_1, ..., s_J$ and proposes a corresponding change in the baseline hazard for the historical and control dataset. Sample $s^* \sim U(s_{j-1}, s_{j+1})$ for $j=1,...,J-1$ and $s^* \sim U(s_{j-1}, \text{min}(\text{max}(\vec{y}),\text{max}(\vec{y}_0))$ for $j=J$ from a continuous uniform distribution. 

To determine $\pi(s_j|s_0, ...s_{J+1})$ for the prior in the acceptance probability where $j < J$, the time increments are expressed as 
\begin{equation}
0=s_0<t_1<s_1<t_3<s_2<t_5<...t_{2J+1}<s_{J+1}.    
\end{equation}
Fixing  $s_0$ and $s_{J+1}$, there are $2J+1$ variables that can move. There are $J$ even ordered variables and $J+1$ odd ordered variables. We can treat all the random variables $t_i$ for $i=1,..,2J+1$, so that when $i$ is even these are the even ordered statistics when $i$ is odd are the odd ordered statistics.

The prior for $\vec{s}$, the even ordered statistics can be obtained by drawing $s_j\sim \text{U}(s_0, s_{J+1})$ and integrating over the joint distribution
\begin{align*}
    \pi(s_1,...,s_J) = \int \dots \int (2J+1)!\frac{1}{(s_{J+1}-s_0)^{2J+1}}  \mathbb{1}_{\left\{s_0<t_1<s_1<t_3<s_2...t_{2J} < s_{J+1}\right\}} dt_1,...,dt_{2J}. 
\end{align*}
Using the integral of $\mathbb{1}_{ \left\{ a<b<c \right\}}$ over $b$ is $1 \times (c-a)$ we end up with 
\begin{align*}
     \pi(s_1,...,s_J) = 
     \frac{(2J+1)!}{(s_{J+1}-s_0)^{2J+1}}\prod_{j=1}^{J+1}s_j-s_{j-1}.
\end{align*}
This allows us to determine the probability of $s_j$  conditioned on all the other values $\bm{s}^{(-j)}$ values. If \newline $s_1, ..., s_{j-1}$, $s_{j+1}$,$...$,$s_{J+1}$ are fixed we have  
\begin{align}
     \label{eq:cond_sj}
     \pi(s_j|\vec{s}^{(-j)}) & = 
     \int \int\frac{1!}{(s_{j+1}-s_{j-1})}  \mathbb{1}_{\left\{s_{j-1}<t_{2j-1}<s_j<t_{2j+1}<s_{j+1}\right\}} dt_{2j-1} dt_{2j+1} \nonumber\\
     & = \frac{(s_{j+1}- s_{j})(s_j- s_{j-1})}{(s_{j+1}-s_{j-1})}\mathbb{1}_{\left\{s_{j-1}<s_j<s_{j+1}\right\}}. 
\end{align}
The probability of accepting, where $j < J$ is 
\begin{align*}
   \text{Ratio} &= 
   \frac{\mathcal{L}(\bm{\beta},\bm{\lambda}, \vec{s}^*, J|\vec{D})
   \mathcal{L}(\bm{\beta}_0,\bm{\lambda}_0, \vec{s}^*, J|\vec{D}_0)\pi(s^*_j|\vec{s}^{(-j)}) (s_{j+1}-s_{j-1})
   }{
    \mathcal{L}(\bm{\beta},\bm{\lambda}, \vec{s}, J|\vec{D})\mathcal{L}(\bm{\beta}_0,\bm{\lambda}_0, \vec{s}, J|\vec{D}_0)\pi(s_j|\vec{s}^{(-j)})(s_{j+1}-s_{j-1})
    }
\end{align*}
and for $j = J$ the conditional prior (\ref{eq:cond_sj}) is adjusted by replacing $s_{j+1}$ with $\text{min}(\text{max}(\vec{y}),\text{max}(\vec{y}_0))$.

\subsection{Reversible-jump moves}

\subsubsection*{Birth move}
By proposing to extend $J$ by 1 we need to propose:
\begin{itemize}
    \item Baseline hazard historical $\lambda_{0j}^*,\lambda_{0j+1}^*$,
    \item Baseline hazard current $\lambda_{j}^*,\lambda_{j+1}^*$,
    \item Between hazard variance $\tau_{j}^*,\tau_{j+1}^*$ ,
    \item Split point $s^*$ and $J^*= J+ 1$. 
\end{itemize}

We sample $s^*$ from a uniform $U_1(0, \text{min}(\text{max}(\vec{y}),\text{max}(\vec{y}_0)))$. Suppose $s^*$, the proposal split lies between the $j$th and $(j-1)$th split times so the ordered times can be expressed as
\begin{align}
\label{eq:time_p}
&(s_0<s_1<s_2<...<s_{j-1}<s^*<s_j<s_{j+1}<...<s_{J+1}) \equiv\\
&(s^*_0<s^*_1<s^*_2<...<s^*_{j-1}<s^*_j<s^*_{j+1}<...<s^*_{J+2}). \nonumber
\end{align}
As in \cite{Green1995} we propose the associated baseline hazard for the interval, for both the historical and current data with a weighted mean on the log scale. The proposed new heights $\lambda_j^*, \lambda_{j+1}^*$  for the step function on the sub-intervals $(s_{j-1}, s^*)$ and $(s_{j}, s^*)$ recognise that the current height $\lambda_j$ on the union of these two intervals is typically well-supported in the posterior distribution and should therefore not be completely discarded. The new heights  $\lambda_j^*, \lambda_{j+1}^*$ are perturbed in either direction from $\lambda_j$ in such a way that $\lambda_j$ is a compromise between them. To preserve positivity and maintain simplicity in the acceptance ratio calculations, a weighted geometric mean is used. This approach also lends itself to the proposal for $\tau^*_j$ and $\tau^*_{j+1}$. 
 
Here we focus on the current baseline hazard, the historical baseline hazard is calculated in the same fashion. 

If we draw $U_2\sim U(0,1)$ and set $\frac{\lambda_{j+1}^*}{\lambda_{j}^*} = \frac{1-U_2}{U_2}$ the weighted mean on the log scale is 
\begin{equation}
    (s^*-s_{j-1})\log(\lambda_j^*) + (s_j-s^*)\log(\lambda_{j+1}^*) = (s_j-s_{j-1})\log(\lambda_j)  
\end{equation}
This approach accommodates the constraint within the reversible jump update. We have two updates to make for the baseline hazards when we propose adding a split point $s^*$, but are allowed only 1 proposal random variable in order to maintain the same number of parameters and additional random variables across the  birth and reverse move. 

This gives 
\begin{align*}
    \log(\lambda_j^*) =& \log(\lambda_j) - \frac{(s_j - s^*)}{(s_j -s_{j-1})}\log\left(\frac{1-U_2}{U_2}\right)\\
    \log(\lambda_{j+1}^*) =& \log(\lambda_j) + \frac{(s^* - s_{j-1})}{(s_j -s_{j-1})}\log\left(\frac{1-U_2}{U_2}\right).
\end{align*}
Using a draw from $U_3 \sim (0,1)$ the historical baseline hazard proposal
\begin{align*}
    \log(\lambda_{0j}^*) =& \log(\lambda_{0j}) - \frac{(s_j - s^*)}{(s_j -s_{j-1})}\log\left(\frac{1-U_3}{U_3}\right)\\
    \log(\lambda_{j+1}^*) =& \log(\lambda_j) + \frac{(s^* - s_{j-1})}{(s_j -s_{j-1})}\log\left(\frac{1-U_3}{U_3}\right). 
\end{align*}
And finally, if $\tau$ which controls the extent of borrowing takes a piecewise form (is indexed by $j$),  $U_4 \sim U(0,1)$ and $\frac{\tau^*_{j+1}}{\tau^*_{j}} = \frac{1-U_4}{U_4}$ is thus
\begin{align*}
    \log(\tau_j^*) =& \log(\tau_j) - \frac{(s_j - s^*)}{(s_j -s_{j-1})}\log\left(\frac{1-U_4}{U_4}\right)\\
    \log(\tau_{j+1}^*) =& \log(\tau_j) + \frac{(s^* - s_{j-1})}{(s_j -s_{j-1})}\log\left(\frac{1-U_4}{U_4}\right).
\end{align*}

In the case of $\vec{s}^*$ the prior, using the left hand side of the identity, (\ref{eq:time_p}) is 
\begin{equation}
   \pi(\vec{s}^*) =  \frac{(2(J+1) +1)!}{(s_{J+1}-s_0)^{2(J+1)+1}}(s_1-s_0)(s_2-s_1)...(s^*-s_{j-1})(s_j-s^*)...(s_{J+1}-s_{J}).
\end{equation}

The prior ratio is thus
\begin{align*}
    \frac{
    \pi(J+1|\phi) N_{J+2}(\log(\bm{\lambda}^*)|\bm{\lambda}^*_0,\text{diag}(\bm{\tau}^*))
    N_{J+2}(\bm{\lambda}^*_0| \mu\vec{1}, \sigma^2\Sigma_{s^*})
    }{
    \pi(J|\phi) N_{J+1}(\log(\bm{\lambda})|\bm{\lambda}_0,\text{diag}(\bm{\tau}))
    N_{J+1}(\bm{\lambda}_0| \mu\vec{1}, \sigma^2\Sigma_{s})
    } \times \\
    \frac{
    \pi(\tau_j^*|\vec{s})\pi(\tau_{j+1}^*|\vec{s}) (s^*-s_{j-1})(s_j-s^*)(2J+3)(2J+2)
    }{
    \pi(\tau_j|\vec{s})(s_j-s_{j-1})(s_{J+1})^2
    }.
\end{align*}

Defining the probability of a birth as $\pi_b$ and death as $\pi_d$ the proposal ratio is thus
\begin{equation*}
    \text{proposal ratio} = 
    \frac{
    \pi_d \frac{1}{J+1}
    }{
    \pi_b\frac{1}{(\text{min}(\text{max}(\vec{y}),\text{max}(\vec{y}_0)) -s_0)}\frac{1}{(1-0)}\frac{1}{(1-0)}
    },
\end{equation*}
where the death moves involves selection one of the $J+1$ split points to kill.

If we define the variables for the current position and the proposal random variables as 
\begin{align*}
    &s_1,s_2,...,s_j,U_1, s_{j+1},...s_{J+1}, \lambda_1, ...,\lambda_j,U_2, \lambda_{j+1},..., \lambda_{J+1},\\
    &\lambda_{01}, ...,\lambda_{0j},U_3, \lambda_{0j+1},..., \lambda_{0J+1},\tau_1, ...,\tau_j,U_4, \tau_{j+1},..., \tau_{J+1}
\end{align*}
where $U_1$ is a uniform random variable from $U(0,\text{min}(\text{max}(\vec{y}),\text{max}(\vec{y}_0)))$ and $U_2$, $U_3$ and $U_4$ are from $U(0,1)$. The proposed parameters are thus  
\begin{equation}
    s_1,s_2,...,s_{j-1},s^*, s_{j},...s_{J+1}, \lambda_1^*, ...,\lambda_{J+2}^*, \lambda^*_{01}, ..., \lambda^*_{0J+2}, \tau_1^*, ...,\tau_{J+2}^*.
\end{equation}
The Jacobian reduces to a product of three determinants
\begin{align*}
    \text{Jacobian}  &=
    \begin{vmatrix}
        \frac{d\lambda_{j}^*}{d\lambda_{j}} & \frac{d\lambda_{j}^*}{dU_2}\\
        \frac{d\lambda_{j+1}^*}{d\lambda_{j+1}} & \frac{d\lambda_{j+1}^*}{dU_2} 
    \end{vmatrix} \times 
    \begin{vmatrix}
        \frac{d\lambda_{0j}^*}{d\lambda_{0j}} & \frac{d\lambda_{0j}^*}{dU_3}\\
        \frac{d\lambda_{0j+1}^*}{d\lambda_{0j+1}} & \frac{d\lambda_{0j+1}^*}{dU_3} 
    \end{vmatrix} \times 
    \begin{vmatrix}
        \frac{d\tau_{j}^*}{d\tau_{j}} & \frac{d\tau_{j}^*}{dU_4}\\
        \frac{d\tau_{j+1}^*}{d\tau_{j+1}} & \frac{d\tau_{j+1}^*}{dU_4} 
    \end{vmatrix} \\
    &= \frac{1}{U_2(1-U_2)} \frac{1}{U_3(1-U_3)} \frac{1}{U_4(1-U_4)}. 
\end{align*}

The probability of accepting a birth move is:
\begin{align*}
\text{Likelihood} =& 
\frac{L(\bm{\beta_0}, J+1, \vec{s}^*, \log(\bm{\lambda}^*_0)|\vec{D}_0)}{L(\bm{\beta_0}, J, \vec{s}, \log(\bm{\lambda}_0)|\vec{D}_0)}  
\frac{L(\bm{\beta}, J+1, \vec{s}^*, \log(\bm{\lambda}^*)|\vec{D})}{L(\bm{\beta}, J, \vec{s}, \log(\bm{\lambda})|\vec{D})} \\
\\
\text{Prior}  = &
     \frac{
    \pi(J+1|\phi) N_{J+2}(\log(\bm{\lambda}^*)|\bm{\lambda}^*_0,\text{diag}(\bm{\tau}^*))
    N_{J+2}(\bm{\lambda}^*_0| \mu\vec{1}, \sigma^2\Sigma_{s^*})
    }{
    \pi(J|\phi) N_{J+1}(\log(\bm{\lambda})|\bm{\lambda}_0,\text{diag}(\bm{\tau}))
    N_{J+1}(\bm{\lambda}_0| \mu\vec{1}, \sigma^2\Sigma_{s})
    } \times \\
    &\frac{
    \pi(\tau_j^*|\vec{s})\pi(\tau_{j+1}^*|\vec{s}) (s^*-s_{j-1})(s_j-s^*)(2J+3)(2J+2)
    }{
    \pi(\tau_j|\vec{s})(s_j-s_{j-1})(s_{J+1})^2
    }.\\
    \\
\text{Proposal} = &
    \frac{
    \pi_d \frac{1}{J+1}
    }{
    \pi_b\frac{1}{(\text{min}(\text{max}(\vec{y}),\text{max}(\vec{y}_0)) -s_0)}\frac{1}{(1-0)}\frac{1}{(1-0)}\frac{1}{(1-0)}
    }\\
    \\
\text{Jacobian} = &
\frac{1}{U_2(1-U_2)} 
\frac{1}{U_3(1-U_3)} 
\frac{1}{U_4(1-U_4)} 
\end{align*}

\subsubsection*{Death move}
The acceptance probability for the corresponding reverse move has the same form with
the appropriate change of labelling of the partitions and variables, and the ratio terms
inverted. First, we sample one of the  split times via a uniform discrete distribution, $s_j \sim \text{min}(\text{max}(\vec{y}),\text{max}(\vec{y}_0))$. The proposal partition of time
axis consists the time splits as follows:
\begin{align}
\label{eq:time_p_d}
&(s_0<s_1<s_2<...<s_{j-1}<s_{j+1}<...<s_{J+1}) \equiv\\
&(s^*_0<s^*_1<s^*_2<...<s^*_{j-1}<s^*_j<s^*_{j+1}<...<s^*_{J}). \nonumber
\end{align}

The log baseline hazard from the current data is 
\begin{equation}
    (s_j-s_{j-1})\log(\lambda_j) + (s_{j+1}-s_j)\log(\lambda_{j+1}) = (s_{j+1}-s_{j-1})\log(\lambda_j^*)  
\end{equation}
Using the perturbation from a uniform distribution for the reverse move
\begin{equation}
    \frac{\lambda_{j+1}}{\lambda_j} = \frac{1-U^*_1}{U^*_1}
\end{equation}

The prior for $\vec{s}^*$ is 
\begin{equation}
   \pi(\vec{s}^*) =  \frac{(2(J-1) +1)!}{(s_{J+1}-s_0)^{2(J+1)+1}}(s_1-s_0)...(s_{j+1}-s_{j-1})...(s_{J+1}-s_{J})
\end{equation}

The prior ratio is 
\begin{align*}
    \text{Prior}  = &
    \frac{
    \pi(J-1|\phi) N_{J}(\log(\bm{\lambda}^*)|\bm{\lambda}^*_0,\text{diag}(\bm{\tau}^*))
    N_{J}(\bm{\lambda}^*_0| \mu\vec{1}, \sigma^2\Sigma_{s^*})
    }
    {
    \pi(J|\phi) N_{J+1}(\log(\bm{\lambda})|\bm{\lambda}_0,\text{diag}(\bm{\tau}))
    N_{J+1}(\bm{\lambda}_0| \mu\vec{1}, \sigma^2\Sigma_s)
    } \times \\
    &\frac{
    \pi(\tau^*_j|\vec{s})s_{J+1}^2(s_{j+1}-s_{j-1})
    }{
    \pi(\tau_j|\vec{s})\pi(\tau_{j+1}|\vec{s})(s_j-s_{j-1})(s_{j+1}-s_j)(2J+1)(2J)
    }\\
    \\
\end{align*}
Defining the probability of a birth as $\pi_b$ and death as $\pi_d$ the proposal ratio is thus
\begin{equation*}
    \text{proposal ratio} = 
    \frac{
    \pi_b\frac{1}{(\text{min}(\text{max}(\vec{y}),\text{max}(\vec{y}_0)) -s_0)}\frac{1}{(1-0)}\frac{1}{(1-0)}\frac{1}{(1-0)}
    }{
    \pi_d \frac{1}{J}
    }.
\end{equation*}

The Jacobian reduces to a product of determinants 
\begin{align*}
    \text{Jacobian} &=
    \begin{vmatrix}
        \frac{d\lambda_{j}}{d\lambda_{j}^*} & \frac{d\lambda_{j}}{d\lambda_{j+1}^*}\\
        \frac{d U_2}{d\lambda_{j}^*} & \frac{d U_2}{d\lambda_{j+1}^*} 
    \end{vmatrix} \times 
    \begin{vmatrix}
        \frac{d\lambda_{0j}}{d\lambda_{0j}^*} & \frac{d\lambda_{0j}}{d\lambda_{0j+1}^*}\\
        \frac{d U_3}{d\lambda_{0j}^*} & \frac{d U_3}{d\lambda_{0j+1}^*} 
    \end{vmatrix} \times
    \begin{vmatrix}
        \frac{d\tau_{j}}{d\tau_{j}^*} & \frac{d\tau_{0j}}{d\tau_{0j+1}^*}\\
        \frac{d U_4}{d\tau_{j}^*} & \frac{d U_4}{d\tau_{0j+1}^*} 
    \end{vmatrix}\\
    & = U_2(1-U_2)U_3(1-U_3)U_4(1-U_4).     
\end{align*}

\section{Understanding the hyperparameters of the commensurate prior}

\subsection{Prior predictive}

Here we describe the posterior predictive over multiple datasets. The intuition of this result is used to understand the approximate worth of our prior. 

The posterior predictive distribution has an inflated variance compared with the posterior. For example, in a Gaussian hierarchical model with known data variance $\sigma^2_i$ and the between trial variability $\tau$. A closed form for $p(\mu|\bar{x}_1,...,\bar{x}_h)$ is available by deriving the joint posterior of $\theta_1,..,\theta_H$ and $\mu$ and integrating over $\theta_1,...,\theta_H$. 

Defining the following model
\begin{align*}
 \bar{x}|\theta_i \sim N(\theta_i, \sigma^2_i)\\
 \theta_1, \theta_H, \theta^*|\mu \sim N(\mu, \tau)
\end{align*}
and fixing $\tau$ and $\sigma^2$, the joint update is conjugate. The posterior for $\mu$ over the historical data is
\begin{equation}
\mu|\bar{x}_1,...,\bar{x}_H \sim N\left(
\frac{\sum w_i \bar{x}_i}{\sum w_i}, \frac{1}{\sum w_i}
\right)
\end{equation}
where $w_i = 1 / (\sigma^2_i + \tau)$ which are the inverse weights from a classic meta-analysis. The posterior predictive 
\begin{equation*}
    p(\theta^*| \vec{D}) = \int p(\theta^*|\mu)p(\mu| \vec{D}) d\mu,
\end{equation*}
is
\begin{equation*}
      \theta^*| \vec{D} \sim \mathcal{N}\left(
      \frac{\sum w_i \bar{x}_i}{\sum w_i}
     , \frac{1}{\sum w_i} + \tau
      \right).
\end{equation*}

The posterior predictive inflates the posterior distribution of $\mu$ by $\tau$ in the conjugate Gaussian hierarchical model with known variance. Using a posterior predictive adds more uncertainty to the informed part of the prior. 

The update for $\theta_i$ where $i=1,...,H$ is a conditional posterior Gaussian distribution
\begin{equation*}
    \theta_i|\vec{D}, \mu \sim N\left(
    \bar{x}_i \frac{\tau^2}{\sigma_i^2 +\tau} + 
    \mu \frac{\sigma_i^2}{\sigma_i^2 +\tau}, \frac{\tau \sigma_i^2}{\tau + \sigma_i^2}
    \right).
\end{equation*}
The expectation of $\theta_i$ is a weighted average of the observer sample mean for $i$ and the population mean. When $\tau$ is small, the expectation is primarily the population mean $\mu$ (which it self is made up of the weighted average of the data across the groups. When $\tau$ is large the expectation is primarily from the data for the $i$th group only.

\subsection{All commensurate prior}
The commensurate prior for the variance which is fixed across the intervals is 
\begin{equation}\label{eq:tau_all_adx}
    \tau \sim p_0\text{IGamma}(a_\tau, b_\tau) + (1-p_0)\text{IGamma}(c_\tau, d_\tau).
\end{equation}

As in Section 2.3 of the main article, we explore the conditional conjugate update for $\tau$ given the prior (\ref{eq:tau_all_adx}).    

The conjugate posterior update for the $\tau^{(\text{all})}$ prior is 
\begin{align}
    \tau^{(\text{all})}|\cdot  \sim&  (q)  \text{IG}\left(\frac{J + 1}{2} +a_\tau, \sum_j \frac{(\log(\lambda_j)-\log(\lambda_{0j}))^2}{2} +b_\tau\right) +  \nonumber\\
    & (1-q) \text{IG} \left(\frac{J + 1}{2}+c_\tau, \sum_j  \frac{(\log(\lambda_j)-\log(\lambda_{0j}))^2}{2} +d_\tau\right).
\end{align}
Compared with the mixture prior for each interval, the shape parameter is a function of the number of intervals, rather than being increased by half. Increasing the shape parameter increases the height of the peak, decreasing the density in the tail. The scale is now a function of the sum of squared log baseline hazard difference across all intervals, rather than interval-specific, increasing the density in the tail.  

The posterior weight
\begin{align}
w_{0} &= \frac{ b_\tau^{a_\tau}\Gamma(a_\tau + \frac{J + 1}{2})}{(\frac{1}{2}\sum_j (\log(\lambda_j) - \log(\lambda_{0j}))^2)+b_\tau)^{\frac{1}{2}+a_\tau}\Gamma(a_\tau) },\\
w_{1} & =  \frac{ d_\tau^{c_\tau}\Gamma(c_\tau + \frac{J + 1}{2})}{(\frac{1}{2}\sum_j(\log(\lambda_j) - \log(\lambda_{0j}))^2)+d_\tau)^{\frac{1}{2}+c_\tau}\Gamma(c_\tau) }
\end{align}
$w = p_{0}w_{0} + p_{1}w_{1}$, 
$q = \frac{p_0w_{0}}{w}$ and $(1- q) = \frac{p_1w_{1}}{w}$.

\begin{equation}\label{eq:qMSE_all}
 q(\text{SSE}|b_\tau, d_\tau, p_0) = \frac{1}{1 + \frac{(1-p_0)}{p_0}\frac{d_\tau}{b_\tau} \left(
 \frac{\frac{1}{2}\text{SSE}+b_\tau}{\frac{1}{2}\text{SSE}+d_\tau}\right)^{\frac{3}{2}}
 }. 
\end{equation}

\section{Comparator models}

\subsection{Informed Mixture prior}
 With one source of historical data one can create an informative prior via a gamma distribution for each segment of the survival curve. If we treat each baseline hazard as independent   
\begin{align*}
    t_{i} &\sim PE(\lambda_1, ..., \lambda_{J+1}, s_0, ..., s_{J+1}) \text{ $J$ split points.} \\
    \lambda_j &\sim G(a_j , b_j)
\end{align*}
then the posterior of the model for the control data with no covariates from the likelihood described in  (\ref{eq:conc_like}) is 
\begin{equation}
    \lambda_j | \bm{\lambda}^{(-j)} \sim G\left(a_j + d_j, b_j +  \sum_{i} t_{ij} \right)
\end{equation}
where $t_{ij}$ is the total exposure time within the $j$th interval and   $d_j$ is the number of events in interval $(s_{j-1}, s_j]$. To construct an informed prior for the control arm where the total exposure time was 1000 weeks and you have 100 events and one point. Then your prior would be $\lambda_j\sim G(50, 500)$. If you wish to weight the prior by a $1/2$ then you have a $\lambda_j\sim G(25, 250)$. This simply sets the prior hyperparameters for the historic data to 0, assumes equal events per time period and and equal split of exposure time.  

You can parameterise the Gamma prior as 
\begin{equation}
    \lambda_j \sim G\left( 
    \frac{\mu_{\lambda_j}}{c}, \frac{1}{c}
    \right)
\end{equation}
where the hyperparameter $\mu_{\lambda_j}$ is the prior mean and the hyperparameter $c$ quantifies the dispersion

\subsection{Hierarchical borrowing}

A hierarchical model is also used. The advantage of this approach is that the borrowing is now dynamic through the update of the variance parameter $\tau^2$. The modelling only requires the total number of events and the total exposure time for each historical data source.  

For the historical study $h =0$ and current study $ h= 1$ a joint model based on a frailty approach, where the random effect is between trials, is 
\begin{equation*}
\begin{array}{l}
    \lambda_{ijh} = \lambda_j  \exp\left\{\gamma_h + x_i \beta\right\} \\
    \lambda_j \sim \mathcal{G}(a_\lambda , b_\lambda) \\
    \gamma_h \sim \mathcal{N}(\mu_\gamma, \tau^2) \\
    \mu_\gamma \sim \mathcal{N}(0, t^2_\gamma) \\
    \tau^2 \sim \mathcal{IG}(a_\tau,b_\tau)\\
    \beta \propto 1
\end{array}
\end{equation*}
A conditionally conjugate Gamma prior is placed on each piecewise baseline hazard $\lambda_j$. The $\gamma_h$  terms are the relative study-level effects (random effect) on the hazard for segment $s$.  A hierarchical model is posited across these study-level effects with $\gamma_h$ given a normal distribution with hyperpriors $\mu_\gamma$ and $\tau^2$. In this setting we have to set values for both of these hyperpriors where the inverse Gamma prior on $\tau^2$ can be specified in terms of mean and weight.  

The likelihood is 
\begin{align}
    \mathcal{L}(\bm{\beta}, \bm{\lambda}, J, \bm{s}|D) =& \prod_{h=1}^H \prod_{i=1}^{n_h}\prod_{j=1}^{J+1} (\lambda_{j} \exp\left\{\gamma_h + x_i\beta \right\})^{\delta_{ijh}\nu_{ih}}
     \nonumber \exp \Big\{
    -\delta_{ijh} \Big(
    \lambda_j(y_{ih}-s_{j-1})\\
    & \qquad  + \sum_{g=1}^{j-1}\lambda_{g} (s_g-s_{g-1})
    \Big)\exp\{\gamma_h   + x_i \beta \}
    \Big\}
\end{align}
Where $\delta_{ijh} = 1$ if the $i$th subject failed or censored in the $j$th fixed interval in the $h$ study. The covariate $x_i$ indicates treatment, for the historic study this is set to $x_i = 0$. 

This can be expressed at the group level data where $k = 1,2$ defines the control or treatment group respectively, thus $d_{hjk}$ denotes the number of events within an interval and $t_{hjk}$ is the total exposure time within the interval, for study $h$, interval $j$ and treatment group $k$. Thus $d_{0j1}$ is the number of historical events for the control and $d_{1j2}$ is the number of current events for the treatment. 
\begin{align*}
     \log(\mathcal{L}) = & \sum_j \lambda_j(d_{0j1} + \sum_k d_{1jk}) + \gamma_0 d_{0j1} + \sum_k d_{1jk}(\gamma_1 +x_k\beta)  -\\
     &\sum_{h=0}^1\sum_{j =1}^{J+1}\sum_{k=1}^2 (\lambda_j t_{hjk}) \exp\{\gamma_h   + x_k \beta \}
\end{align*}
where $x_2  =1$ and $x_1 = 0$.  

The conditional conjugate updates are the following
\begin{align*}
    \lambda_j|\cdot & \sim G\left(a_{\lambda} + d_j +d_{0j}, b_{\lambda} + \sum_{i=1}^{n_1}\exp\left\{ 
    x_i \beta + \gamma_1
    \right\}t_{1ij} + 
    \sum_{i=1}^{n_0}\exp\left\{ \gamma_0
    \right\}t_{0ij} 
    \right) \\
    \tau^2| \cdot & \sim IG\left( a_\tau+1, b_\tau + \sum_{h=0}^1 \frac{(\gamma_h - \mu_\gamma)^2}{2}\right) \\
    \mu_\gamma|\cdot & \sim N \left(
    \frac{t^2_\gamma}{2t^2_\gamma + \tau^2}(\gamma_0 + \gamma_1), \left(\frac{2}{\tau^2} + \frac{1}{t_\gamma^2}
    \right)^{-1}
    \right)
\end{align*}
which leaves $\gamma_h$ and $\beta$ which are sampled using a Metropolis adjusted Langevin.  

The full log conditional distribution for $\beta$ is 
\begin{align}
    \log p(\beta| \bm{\lambda}, J, \bm{s}, D) \propto & \sum_{i=1}^{n_h}\sum_{j=1}^{J+1}  (\delta_{ij*}\nu_{i*}) x_{i*}\beta  
    -\delta_{ijh} \Big(
    \lambda_j(y_{i*}-s_{j-1})   \nonumber\\
    & \qquad  + \sum_{g=1}^{j-1}\lambda_{g} (s_g-s_{g-1})
    \Big)\exp\{\gamma_*   + x_{i*} \beta \}
\end{align}
Where $\delta_{ijh} = 1$ if the $i$th subject failed or censored in the $j$th fixed interval in the $h$ study. The covariate $x_{ih}$ indicates treatment, for the historic study this is set to $0$ for the historic study.

The first $D_1(\beta)$ derivative of the log conditional posterior are given by 
\begin{align*}
    D_1(\beta) =& \sum_h \sum_i \sum_j (\delta_{ijh}\nu_{ih}) x_{i} + \\
    & - \delta_{ijh}\left(\lambda_{j}(y_{ih}-s_{j-1}) +  \sum_{g=1}^{j-1}\lambda_{g}(s_g -s_{g-1})\right) x_{ih}\exp\left\{
    {x}_{i}\beta + \gamma_h
    \right\}.
\end{align*}
which reduces to 
\begin{align*}
    D_1(\beta) =&  \sum_i \sum_j (\delta_{ij*}\nu_{i*}) x_{i} + \\
    &-\delta_{ij*}\left(\lambda_{j}(y_{i*}-s_{j-1}) +  \sum_{g=1}^{j-1}\lambda_{g}(s_g -s_{g-1})\right) x_{i*}\exp\left\{
    {x}_{i*}\beta + \gamma_*
    \right\}.
\end{align*}
as the covariate is 0 for the historic study. Make the proposal by incorporating the derivative with Gaussian noise as in (\ref{eq:AL_beta}) and then accept with probability equal to the ratios of target and proposal.  

For $\gamma_h$ the first derivative is 
\begin{align*}
    D_1(\gamma_h) =& \sum_i \sum_j (\delta_{ijh}\nu_{ih}) - \delta_{ijh} \left(\lambda_{j}(y_{ih}-s_{j-1}) +  \sum_{g=1}^{j-1}\lambda_{g}(s_g -s_{g-1})\right) \exp\left\{
    {x}_{i}\beta + \gamma_h
    \right\} + \\
    &- 
    \frac{(\gamma_h - \mu_\tau)}{\tau^2},
\end{align*}
which is used to guide the proposal in the sampler.

\section{Sample Size Calculation}
The sample size for our two phase 2 trials is calculated by assuming a target treatment effect, estimating a hazard ratio, and then adjusting the number of required events by the event rate rate. Treatment in the target trial is expected to increase the survival probability at 48 weeks by $40\%$. The survival probability of 0.47 at week 48 is estimated from placebo-controlled patients in the historical data. The hazard ratio, under the proportional hazard assumption, is 
\begin{equation}\label{eq:hr_hist}
    \psi_R = \frac{\log(S(48) \times 1.4)}{\log(S(48))}.
\end{equation}
The target study will assess time-to-first exacerbation during approximately 48 weeks of follow up. The estimated sample size for the phase 2 trial of 84 patients per arm, is calculated by first estimating the required number of events for a two-sided type I error of $10\%$ and a power of $80\%$ to detect a $40\%$ increase in the survival rate, assuming equal allocation. This is then adjusted by the estimated average probability of an event, using the hazard ratio in (\ref{eq:hr_hist}). There is no need to adjust for accrual and follow-up period, as the patients are only lost-follow-up after their last visit for the 48-week treatment period. This is a by product of the primary endpoint of the original phase III trial. 

\begin{figure}[H]
	\centering
	\includegraphics[width = 14cm]{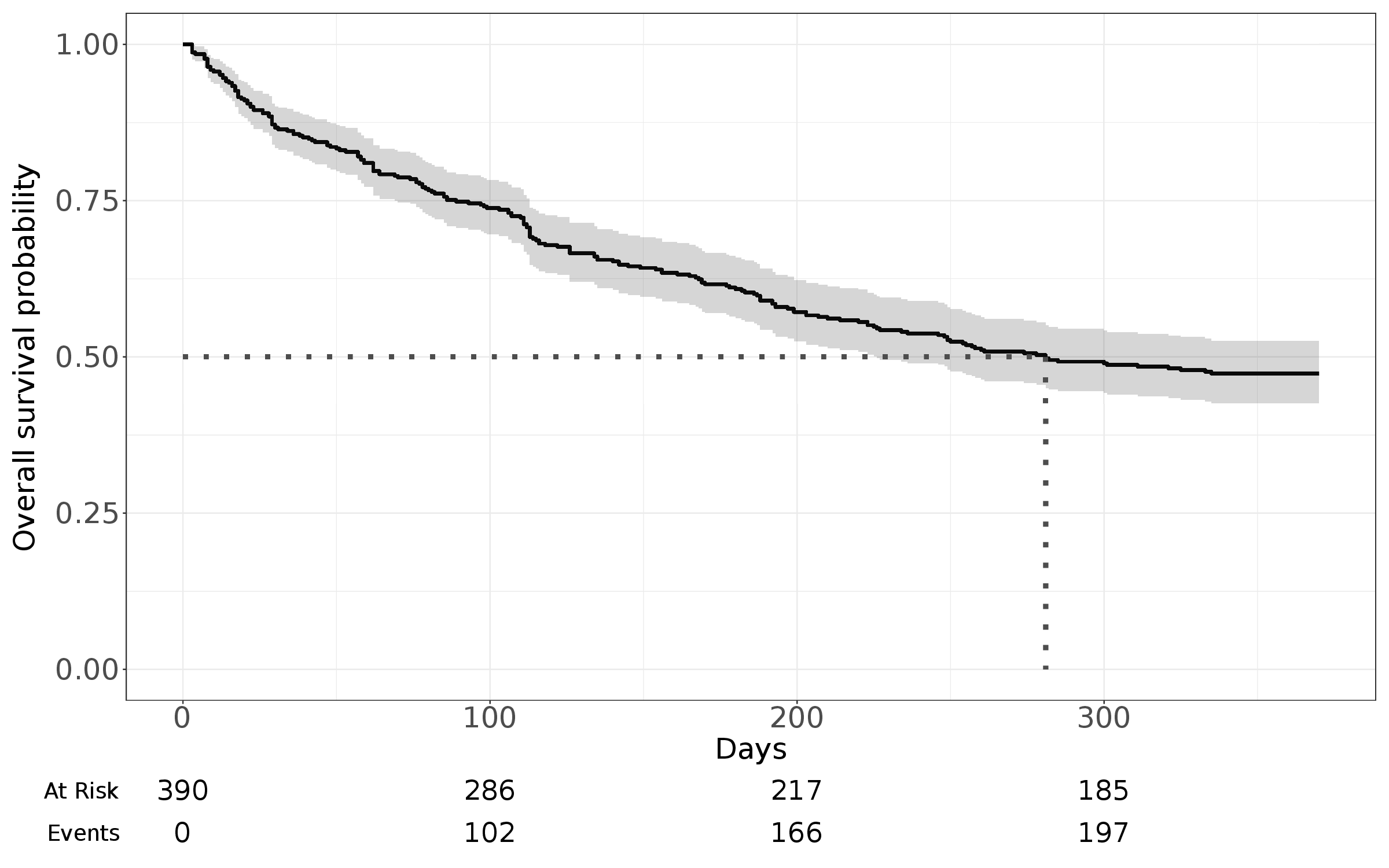}
	\caption{Kaplan-Meier survival curves for the placebo control group from the CALIMA randomized double-blind phase 3 control trial. The survival curve is used to determine the phase 2 sample sizes for the historical and current study.}
	\label{Fig:KM1}
\end{figure}

\section{Additional Figures}
This section contains figures that are referenced in the main paper. 

\begin{figure}[H]
	\centering
	\includegraphics[width=14cm]{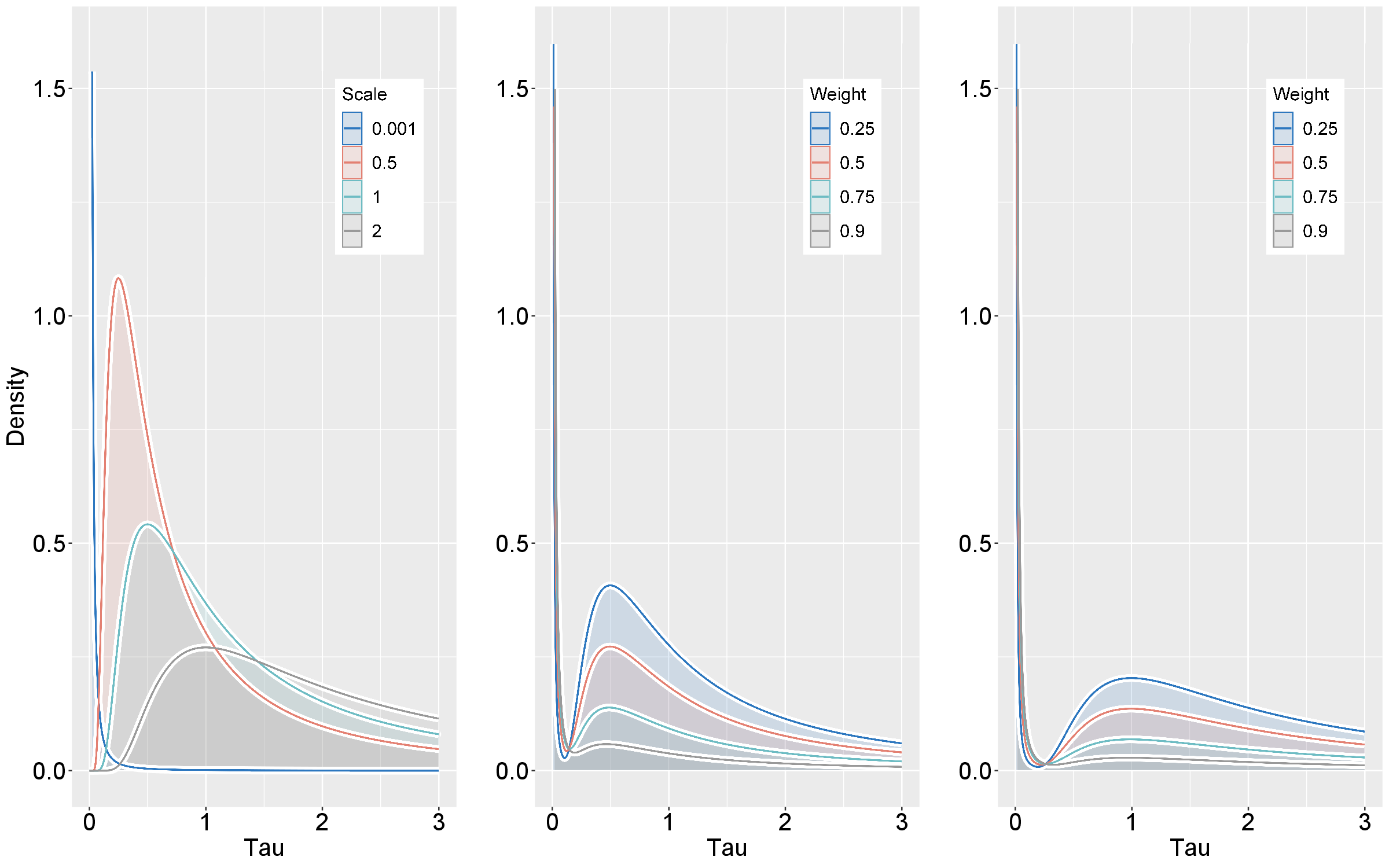}
	\caption{Left panel: Plot of the inverse gamma distribution with shape parameter $a_\tau = 1$ and four different scale parameters $b_\tau$ of 0.001 (blue), 0.5 (red), 1 (green) and 2 (grey). Middle panel: Plot of lump-and-smear inverse gamma priors with shape $a_\tau =1$ , $c_\tau = 1$,  scale parameter $b_\tau = 0.001$,  $d_\tau = 1$ and four different weight parameters $p_0$ of 0.25, 0.5, 0.75, and 0.9. Right panel: Plot of lump-and-smear inverse gamma priors with the same choice of parameters as the middle panel except for a scale parameter of $d_\tau = 2$ rather than $d_\tau = 1$. In all plots, the x and y axes have been truncated to help visual representation.}
	\label{Fig:lumpsmear}
\end{figure}

\begin{figure}[H]
	\centering
	\includegraphics[width=14cm]{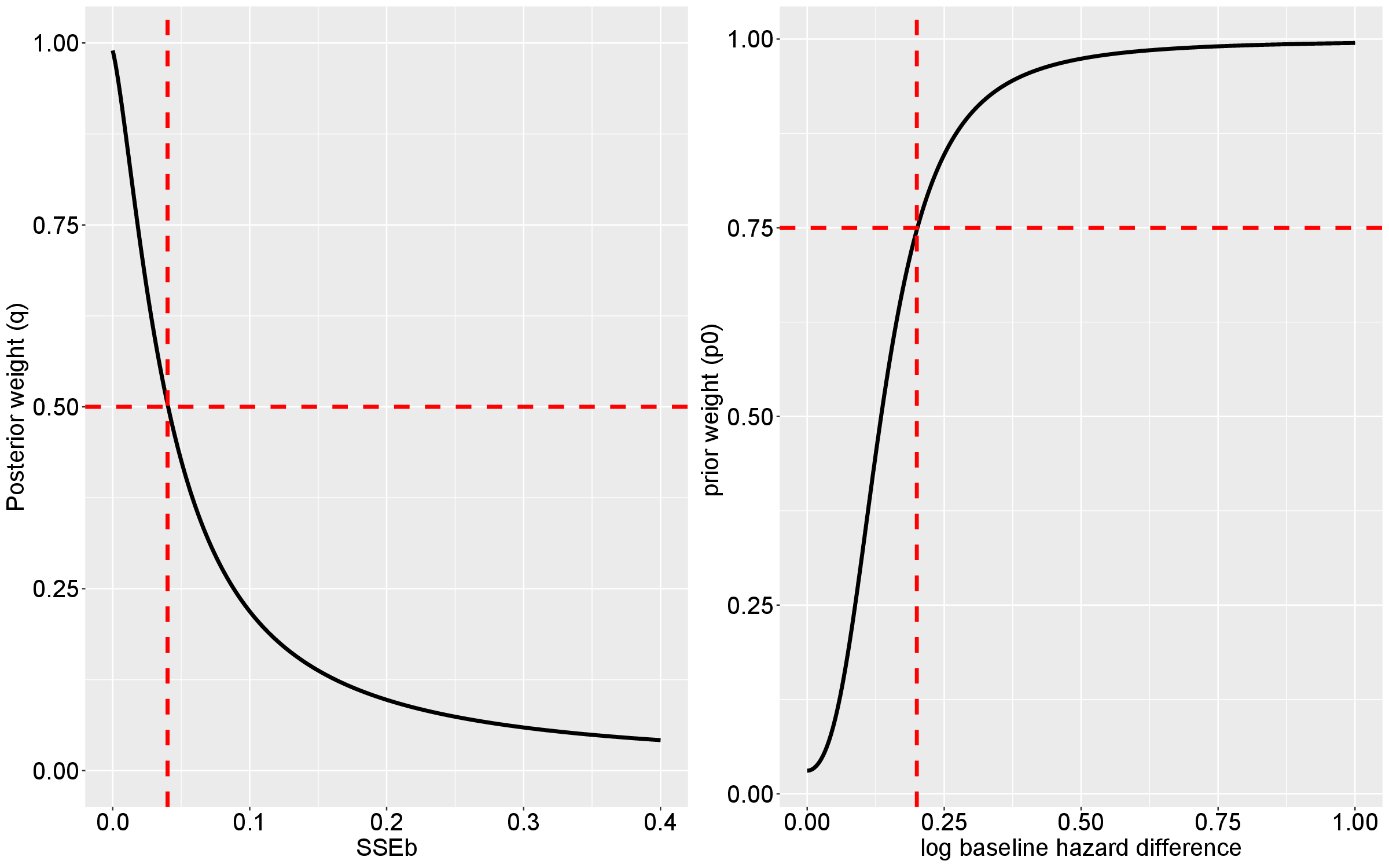}
	\caption{Left panel: Plot of the of posterior weight as a function of the sum of squared error (SSEb) of the log baseline hazards, for hyperparameters of the $\tau^{\text{(mix)}}$ prior $b_\tau = 0.001$, $d_\tau = 1$ and $p_0 =0.75$. The prior weight $p_0$ has been selected so that the tipping-point of $q_j(\text{SSEb}|\cdot)$ is at a 0.2 difference between the log baseline hazards (SSEb = 0.04). Right panel: Plot of the value for $p_0$ from a $\tau^{\text{(mix)}}$ prior $b_\tau = 0.001$, $d_\tau = 1$ for absolute differences between the log baseline hazards. The choice of $p_0=0.75$ for a 0.2 limit of tolerable difference $\xi_j$ is denoted by the red dashed lines. }
	\label{Fig:q_prior}
\end{figure}

\begin{figure}[H]
	\centering
	\includegraphics[width=14cm]{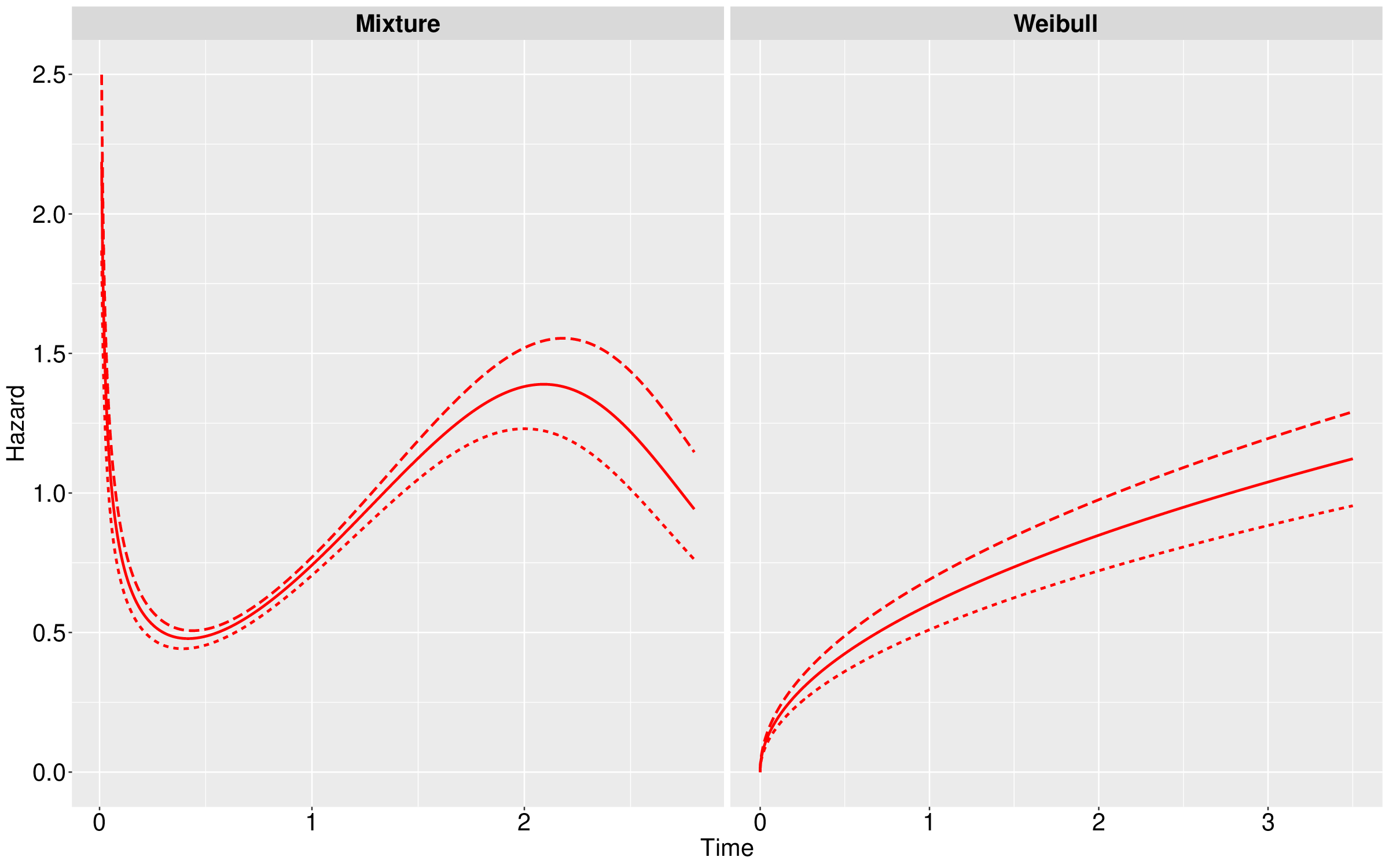}
	\caption{Simulation study: Plot of the true underlying baseline hazards for the simulated  historical datasets from the mixture of Weibull distributions (left) and the Weibull distribution (right). The bold line is for scenario A and B, the long dashed line is for scenario D and the short dashed line is for scenario C.}
	\label{Fig:hazards_sim}
\end{figure}

\begin{figure}[H]
	\centering
	\includegraphics[width=14cm]{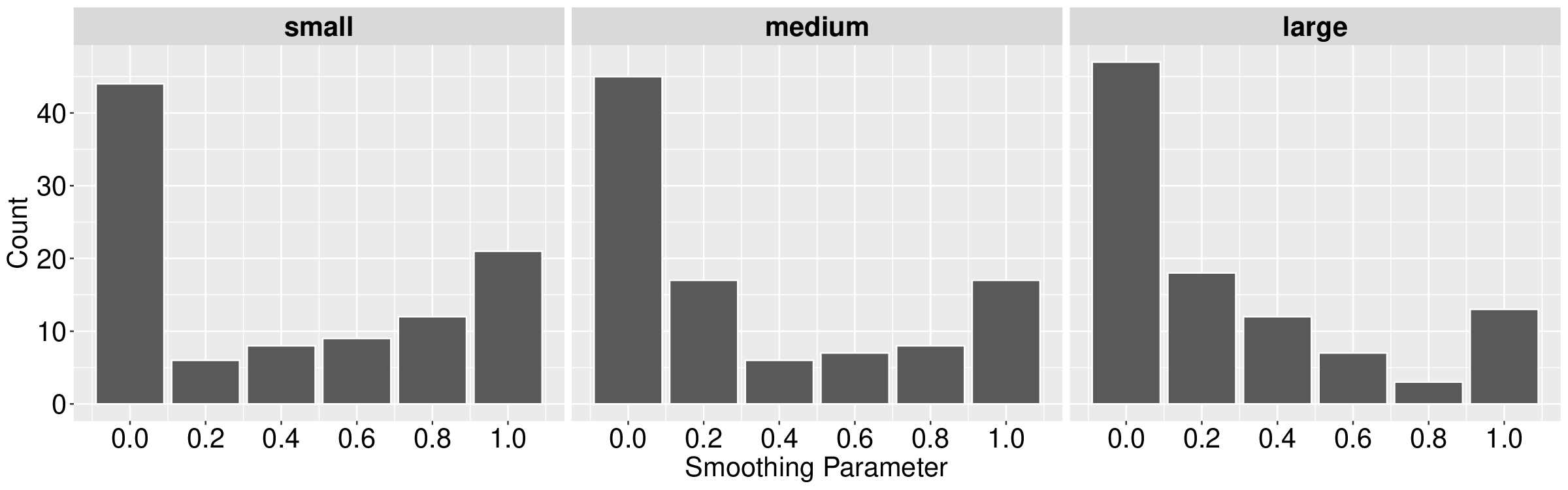}
	\caption{Simulation study current data only: Bar charts displaying the optimal choice of $c_\lambda$, in achieving the smallest MSE for the estimated baseline hazard  for 100 simulated datasets from the Weibull mixture distribution for the three different combinations of $(\phi, J_\text{max})$; small (3, 5), medium (5, 10) and large (10, 20). As the number of allowable split points gets larger and the model becomes more flexible the smoothing parameter is less effective and improving the model fit. }
	\label{Fig:weimix_nb_barcharts}
\end{figure}

\begin{figure}[H]
	\centering
	\includegraphics[width=14cm]{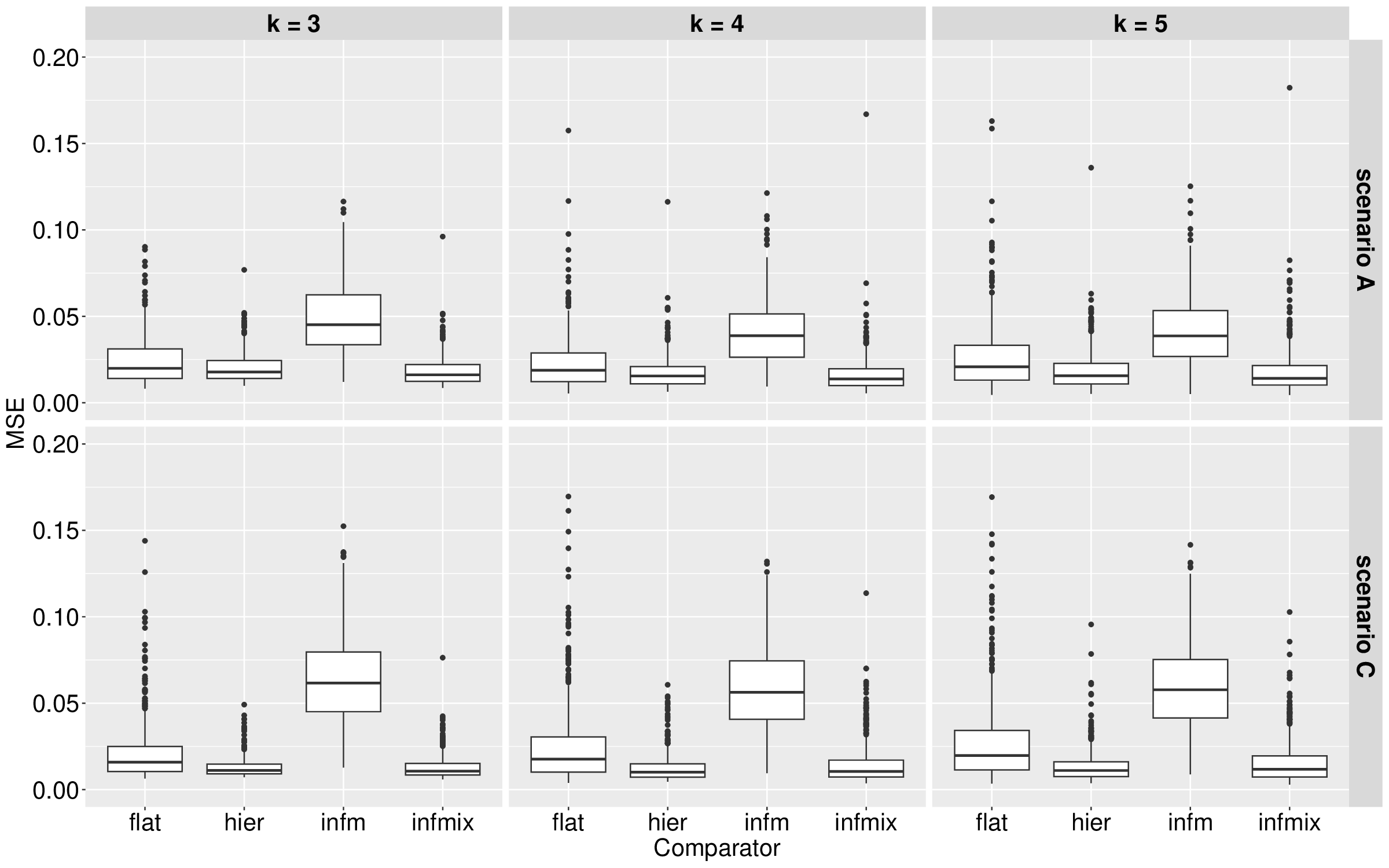}
	\caption{Comparison approaches for scenario A and C: Box plots of the MSE of the current baseline hazard for the comparison approaches  with fixed time intervals at the $100(k/K)^{\text{th}}$ percentile of observed failure times $K = 3,4,5$ for scenarios A and C.}
	\label{Fig:bp_sc13}
\end{figure}

\begin{figure}[H]
	\centering
	\includegraphics[width=14cm]{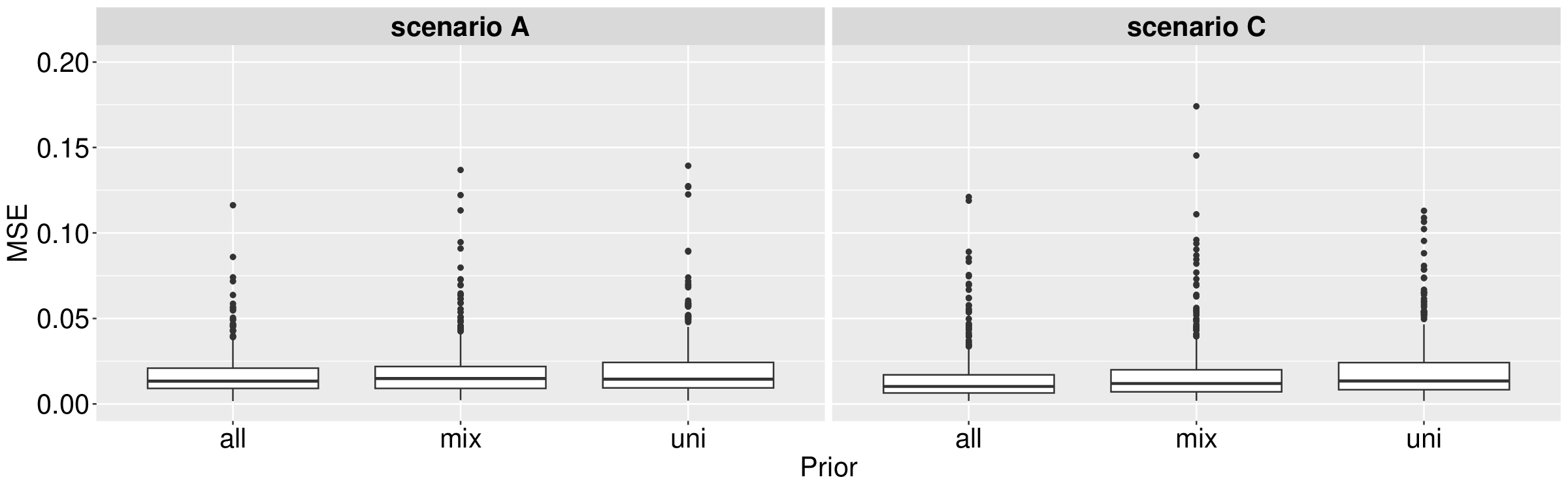}
	\caption{Flexible baseline hazard model for scenario A and C: Box plots of the MSE of the current baseline hazard for the FBHM with the 3 different prior structures all and mix  and uni for scenarios A and C.}
	\label{Fig:ds_bp13}
\end{figure}

\begin{figure}[H]
	\centering
	\includegraphics[width = 14cm]{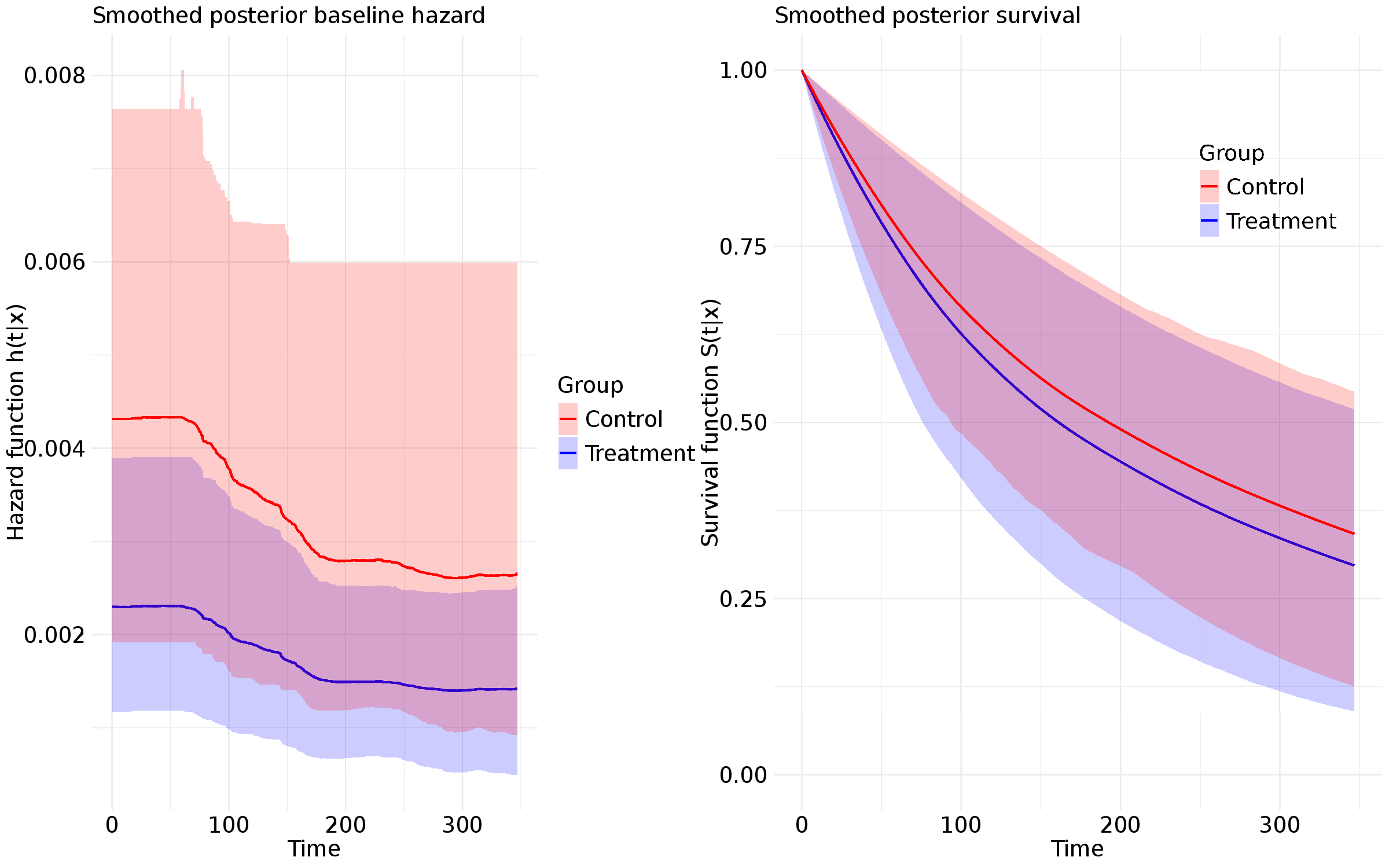}
	\caption{Time is in days. Plots for the model with prior weight of $p_0=0.5$. LHS: Plot of the smoothed posterior baseline hazard for placebo control group and the Q8W treatment group. The red and blue line are the expected posterior hazard, the light blue and red bands are the associated $95\%$ posterior credible intervals. RHS: Plot of the posterior survival function for placebo control group and the Q8W treatment group. The red and blue line are the expected posterior probability of survival, the light blue and red bands are the associated $95\%$ posterior credible intervals. }
	\label{Fig:mcmc_diag}
\end{figure}

\section{Data Analysis Results}
Plots of the smoother posterior and tables of results from the application of our FBHM to the SIRROCO placebo-controlled randomised trial borrowing from external placebo patients from the CALIMA trial. The table includes the $90\%$ highest posterior density credible intervals for each parameter. The region parameter corresponds to the EU, so the expectation for a non-EU patient is -0.031. The effect of 3 exacerbations in the previous year is obtained by combining the posterior samples using the sum-to-zero constraint ($\beta_{\text{e}3} = -\beta_{\text{e}2} - \beta_{\text{e}1} $) and then performing the appropriate inference. 

\begin{table}[H]
    \centering
    \begin{tabular}{lccc}
         \hline \hline
        Parameter&  Mean  & SD & $90\%$ CI \\
         \hline \hline
          Benralizumab Q8W & -0.635  & 0.254 & (-1.049, -0.210) \\
         Region (EU)          & -0.319   & 0.171 & (-0.596, -0.038)\\
         Exacerbation (1)  &  0.100  &  0.176 &  (-0.165, 0.412) \\
         Exacerbations (2)  & 0.098  & 0.207 &  (-0.242,  0.438) \\
         Exacerbations (3)  & -0.198  & 0.242 &  (-0.592,  0.202) \\
         Oral corticosteroids       & 0.772  & 0.164 &  (0.516, 1.049)\\
         \hline
    \end{tabular}
    \caption{Posterior mean and $90\%$ highest posterior density credible intervals from the FBHM for for the prior mixture weight $p_0 = 0.8$.}
    \label{tab:results_da1}
\end{table}

\begin{table}[H]
    \centering
    \begin{tabular}{lccc}
         \hline \hline
        Parameter&  Mean  & SD & $90\%$ CI \\
         \hline \hline
          Benralizumab Q8W & -0.611 & 0.278 & (-1.045, -0.156) \\
         Region (EU)         & -0.333   & 0.186 & (-0.637, -0.032)\\
         Exacerbation(1)  & 0.110  &  0.186 &  (-0.196, 0.406) \\
         Exacerbations (2)  & 0.100  & 0.208 &  (-0.253,  0.431) \\
         Exacerbations (3)  & -0.210  & 0.247 &  (-0.598,  0.203) \\
         Oral corticosteroids          & 0.766 & 0.176 &  (0.487,  1.067)\\
         \hline
    \end{tabular}
    \caption{Posterior mean and $90\%$ highest posterior density credible intervals from the FBHM for for the prior mixture weight $p_0 = 0.5$.}
    \label{tab:results_da2}
\end{table}

\end{refsection}

\end{document}